%% file: 00_main.tex
\documentclass[conference]{IEEEtran}
\IEEEoverridecommandlockouts
\pdfoutput=1
\usepackage{silence} 
\usepackage{cite}
\usepackage{amsmath,amssymb,amsfonts}
\usepackage{algorithmic}
\usepackage{graphicx}
\usepackage{textcomp}
\usepackage{xcolor}
\usepackage{comment}
\usepackage{nohyperref}
\usepackage{xurl}
\usepackage{multirow}
\usepackage{siunitx}
\usepackage{float}
\usepackage{enumitem}
\usepackage{blindtext}
\usepackage{pgfplots}
\pgfplotsset{compat=1.8}
\usepgfplotslibrary{dateplot}
\usepackage{dblfloatfix}

 \newcommand{\red}[1]{\textcolor{red}{#1}}
\makeatletter
\newcommand*\short[1]{\expandafter\@gobbletwo\number\numexpr#1\relax}
\makeatother

\definecolor{color1}{RGB}{0,0,0} 
\definecolor{color2}{RGB}{182,193,251} 
\definecolor{bblue}{HTML}{4F81BD}
\definecolor{rred}{HTML}{C0504D}
\definecolor{ggreen}{HTML}{9BBB59}
\definecolor{ppurple}{HTML}{9F4C7C}
\definecolor{sandybrown}{HTML}{f4a460}
\definecolor{lightseagreen}{HTML}{20b2aa}
\definecolor{cornflowerblue}{HTML}{6495ed}
\definecolor{limegreen}{HTML}{32CD32}
\definecolor{orange}{HTML}{ffa500}
\definecolor{purple}{HTML}{6495ed}

\usepackage{pgfplots}

\usepackage{tikz}
\tikzstyle{Scaling} = [rectangle, rounded corners, minimum width=1.9cm, minimum height=1.5cm,text centered, draw=black, fill=cornflowerblue!30]

\tikzstyle{PCA} = [rectangle, rounded corners, minimum height=1.5cm,text centered, draw=black, fill=sandybrown!30]

\tikzstyle{Undersampling} = [rectangle, rounded corners, minimum height=1.5cm,text centered, draw=black, fill=ppurple!30]

\tikzstyle{Oversampling} = [rectangle, rounded corners, minimum height=1.5cm,text centered, draw=black, fill=lightseagreen!30]

\tikzstyle{Classification} = [rectangle, rounded corners, minimum height=1.5cm,text centered, draw=black, fill=rred!30]

\tikzstyle{Filtering} = [rectangle, rounded corners, minimum width=1.9cm, minimum height=1.5cm,text centered, draw=black, dashed, fill=limegreen!20]

\tikzstyle{Frame} = [rectangle, minimum width=2cm, minimum height=1.5cm,text centered, draw=black, fill=none]

\tikzstyle{arrow} = [thick,->,>=stealth]

\tikzstyle{arrow_small} = [thick,densely dotted,->,>=stealth]

\usepackage{pgfplots}
    \usetikzlibrary{
        pgfplots.dateplot,
    }
\usetikzlibrary{matrix,calc}
\usetikzlibrary{calc}
\usetikzlibrary{positioning}
\usetikzlibrary{shapes.geometric, arrows}

\usepackage{varwidth}

\usepackage[acronym, automake]{glossaries}
\makeglossaries
\input{glossary.tex}
\def\BibTeX{{\rm B\kern-.05em{\sc i\kern-.025em b}\kern-.08em
    T\kern-.1667em\lower.7ex\hbox{E}\kern-.125emX}}

\def\checkSameDate#1{%
    \ifnum\coordindex=0
        \gdef\itsDate{#1}%
    \else
        \def\temp{#1}%
        \ifx\temp\itsDate
        \else
            \PackageError{custom}{Sorry, expected the same date but got two different ones \itsDate\space and #1}{}%
        \fi
    \fi
}%

\def\ensuretwodigits#1{%
    \ifnum#1<10
        \edef\pgfmathresult{0#1}%
    \else
        \edef\pgfmathresult{#1}%
    \fi
}%

\pgfplotsset{
    @datetime to number/.code args={#1-#2-#3 #4:#5:#6}{%
        \checkSameDate{#1-#2-#3}%
        \pgfmathparse{#4*3600+60*#5 + #6}%
    },
    date same day in x/.style={%
        x coord trafo/.style={/pgfplots/@datetime to number=##1},
        scaled x ticks=false,
        plot coordinates/math parser=false,
        x coord inv trafo/.code={%
            \begingroup
            \pgfkeys{/pgf/fpu}%
            \pgfmathparse{mod(##1,60)}%
            \pgfmathfloattoint{\pgfmathresult}%
            \ensuretwodigits{\pgfmathresult}%
            \global\let\seconds=\pgfmathresult
            \pgfmathparse{mod( (##1 - \seconds)/60,60)}%
            \pgfmathfloattoint{\pgfmathresult}%
            \ensuretwodigits{\pgfmathresult}%
            \global\let\minutes=\pgfmathresult
            \pgfmathparse{(##1 - 60*\minutes - \seconds)/3600}%
            \pgfmathfloattoint{\pgfmathresult}%
            \ensuretwodigits{\pgfmathresult}%
            \global\let\hours=\pgfmathresult
            \endgroup
        },
        xticklabel style={
            /pgf/number format/fixed,
        },
        xticklabel=\hours:\minutes:\seconds,
    },
}

\begin{document}

\title{Assessment of Cyber-Physical Intrusion Detection and Classification for Industrial Control Systems


\thanks{This work is funded by Innovation Fund Denmark (File No. 91363) and Horizon 2020 project INTERPRETER under grant agreement No. 864360.}
}

\author{\IEEEauthorblockN{Nils Müller, Charalampos Ziras, Kai Heussen}
\IEEEauthorblockA{\textit{Wind and Energy Systems Department} \\
\textit{Technical University of Denmark}\\
Lyngby, Denmark \\
\{nilmu; chazi; kh\}@dtu.dk}

}

\IEEEoverridecommandlockouts


\maketitle

\begin{abstract}
The increasing interaction of \glspl{ICS} with public networks and digital devices introduces new cyber threats to power systems and other critical infrastructure. 
Recent cyber-physical attacks such as Stuxnet and Irongate revealed unexpected \gls{ICS} vulnerabilities and a need for improved security measures. 
Intrusion detection systems constitute a key security technology, which typically monitors cyber network data for detecting malicious activities.
However, a central characteristic of modern \glspl{ICS} is the increasing interdependency of physical and cyber network processes. 
Thus, the integration of network and physical process data is seen as a promising approach to improve predictability in real-time intrusion detection for \glspl{ICS} by accounting for physical constraints and underlying process patterns. 
This work systematically assesses machine learning-based cyber-physical intrusion detection and multi-class classification through a comparison to its purely network data-based counterpart and evaluation of misclassifications and detection delay. 
Multiple supervised detection and classification pipelines are applied on a recent cyber-physical dataset, which describes various cyber attacks and physical faults on a generic \gls{ICS}.
A key finding is that the integration of physical process data improves detection and classification of all considered attack types.
In addition, it enables simultaneous processing of attacks and faults, paving the way for holistic cross-domain root cause identification.

\end{abstract}


\glsresetall
\input{01_Introduction}

\input{02_Dataset_description_and_evaluation}
\input{03_Methodology}

\input{04_Results}
\input{05_Conclusion}

\bibliographystyle{ieeetr}
\bibliography{References}

\end{document}

%% file: Glossary.tex
\newacronym{ICS}{ICS}{industrial control system}
\newacronym{PLC}{PLC}{programmable logic controller}
\newacronym{HMI}{HMI}{human-machine interface}
\newacronym{IDS}{IDS}{intrusion detection system}
\newacronym{ML}{ML}{machine learning}
\newacronym{CPU}{CPU}{central processing unit}
\newacronym{SCADA}{SCADA}{supervisory control and data acquisition}
\newacronym{SAS}{SAS}{substation automation system}
\newacronym{DSO}{DSO}{distribution system operator}
\newacronym{DoS}{DoS}{denial-of-service}
\newacronym{MiTM}{MITM}{man-in-the-middle}
\newacronym{SVM}{SVM}{support vector machine}
\newacronym{ANN}{ANN}{artificial neural network}
\newacronym{RF}{RF}{random forest}
\newacronym{KNN}{KNN}{k-nearest neighbors}
\newacronym{GB}{GB}{gradient boosting}
\newacronym{SMOTE}{SMOTE}{synthetic minority oversampling technique}
\newacronym{PCA}{PCA}{principal component analysis}
\newacronym{IHT}{IHT}{instance hardness threshold}
\newacronym{CV}{CV}{cross-validation}

%% file: 01_Introduction.tex
\section{Introduction}
In recent years, \glspl{ICS} face an ongoing opening to the internet \cite{asghar2019cybersecurity}. 
Previously isolated systems relying on private networks and specifically designed protocols increasingly use public networks and digital devices to achieve several business benefits.
However, along with advantages such as cost-efficiency and process flexibility, new cyber vulnerabilities emerge, as evidenced by a growing number of cyber attacks on \glspl{ICS} \cite{ThoughtLab2022}.
Besides the introduction of new cyber threats, the ongoing integration of cyber and physical components transforms modern \glspl{ICS} to complex cyber-physical systems. 
In such cyber-physical \glspl{ICS}, failures can originate from a variety of hard- or software faults, human errors or malicious activities. 
Distinguishing attacks from physical faults or human errors is a particularly challenging task in such highly integrated and complex systems, which complicates the identification of root causes. \IEEEpubidadjcol
\vspace{-0.47mm}

\Glspl{IDS} are responsible for detecting malicious activities by monitoring and analyzing either \glspl{ICS} end-device (host-based \gls{IDS}) or network data (network-based \gls{IDS}).
In recent years, the use of \gls{ML} for \glspl{IDS} has attracted increasing interest for reasons such as the ability to capture complex properties of \gls{ICS} operation and attacks, lower \gls{CPU} load compared to conventional \glspl{IDS}, higher detection speed, reduced need for expert knowledge due to generalizability, and the exploitation of steadily increasing amounts of data in \glspl{ICS} \cite{8386762}. 
In this context, supervised multi-class classification comes with the advantage of enabling automated distinction between different types of attacks and other anomalies such as physical faults, facilitating the identification of root causes.

A promising approach to improve supervised intrusion detection and classification is the integration of cyber network with physical process data \cite{ayodeji2020new}.
In this way, the underlying physical constraints and patterns of an \gls{ICS} are included into the \gls{IDS}, potentially improving predictability. 
A few studies investigate such supervised cyber-physical detection and classification of cyber attacks.
In \cite{zhang2019multilayer} the authors propose a multi-layer cyber attack detection system, which combines a supervised and exclusively network data-based classification step with an empirical model for detecting abnormal operation in physical process data. 
However, the authors consider a binary classification problem (normal vs. attack) which weakens the determination of causational factors.
Moreover, as the dataset was shuffled and randomly divided into training and test data, samples of a specific attack event can be found in both sets, which entails data leakage, weakening the validity of the results.
In \cite{yeckle2017evaluation, keshk2017privacy, hink2014machine} \gls{ML}-based intrusion detection and classification in \glspl{ICS} is investigated considering multiple attack types as well as cyber network and physical process features. 
However, none of these works compares cyber-physical with network intrusion detection, or evaluates misclassifications and detection delay.
In \cite{loukas2017cloud, vuong2015performance, vuong2015decision} the benefit of integrating physical process data into a supervised classification-based \gls{IDS} is demonstrated for robotic vehicles. 
These works neither consider a multi-class classification problem nor a scenario including both cyber attacks and physical faults. 
The thematically and methodologically closest works are \cite{faramondi2021hardware} and \cite{sahu2021multi}.
In \cite{faramondi2021hardware} the authors introduce the dataset which also forms foundation for the present work.
To demonstrate a use case of the dataset, they compare several supervised classifiers for intrusion detection.
Although the work compares the use of cyber network and physical process data, the integration of both information sources is not considered.
Moreover, only a binary classification problem is examined.
Finally, models are evaluated using K-fold \gls{CV}, which leads to data leakage as samples from the same attack or fault event are placed in the training and test datasets.
In \cite{sahu2021multi} the authors compare several unsupervised and supervised models for cyber-physical intrusion detection in power systems.
The study explicitly compares network to cyber-physical intrusion detection.
However, in total only four \gls{MiTM} attack events are considered and investigated individually, again leading to data leakage.

The review reveals that existing works on supervised cyber-physical intrusion detection and classification either i) lack systematic assessment through comparison to a purely network data-based approach and evaluation of misclassifications and detection delay, or ii) do not consider multi-class classification, weakening root cause identification, or iii) suffer from methodological issues, limiting the validity of results.

This work systematically assesses real-time cyber-physical intrusion detection and multi-class classification based on a comparison to its exclusively network data-based counterpart and evaluation of misclassifications and detection delay.
Various supervised detection and classification pipelines are implemented and evaluated on a recent dataset of a generic \gls{ICS} \cite{faramondi2021hardware}, describing several cyber attack and physical fault types based on physical process and cyber network data.

\subsection{Contribution and paper structure}
The main contributions of this work are as follows:
\begin{itemize}
    \item Systematic comparison of \gls{ML}-based network and cyber-physical intrusion detection and multi-class classification for \glspl{ICS}, evaluating multiple classification pipelines.
    \item Attack and fault class-wise analysis of misclassifications and detection delay for cyber-physical intrusion detection.
    \item Proposal and assessment of prediction filtering for reduction of misclassifications. 
    \item Transferability evaluation of the investigated generic \gls{ICS} to control systems in the power sector. 
\end{itemize}

The remainder of the paper is structured as follows. In Section \ref{sec:dataset_and_transferability} the investigated dataset is introduced, and the transferability of the underlying \gls{ICS} to power sector control systems evaluated. 
Section \ref{sec:methodology} provides a description of the data preparation as well as applied models and techniques. 
In Section \ref{sec:evaluation} results are presented and discussed, followed by a conclusion and a view on future work in Section \ref{sec:conclusion}.

%% file: 02_Dataset_description_and_evaluation.tex
\section{Dataset description and transferability} \label{sec:dataset_and_transferability}
This section first describes the dataset under investigation (Subsection \ref{sec:dataset}). 
Thereafter, Subsection \ref{sec:transferability} sheds light on the transferability of this work's results by comparing the investigated generic \gls{ICS} to control systems in the power sector.

\subsection{Dataset} \label{sec:dataset}
The dataset used in this study was acquired from a hardware-in-the-loop water distribution testbed and is introduced in \cite{faramondi2021hardware}.
The system distributes water across eight tanks, where one process cycle is defined by a full filling/emptying process of each tank. 
This procedure is steadily repeated, rendering it a cyclical process.
Water flow between the tanks is realized by valves, pumps, pressure sensors and flow sensors. 
The process is controlled by a typical \gls{SCADA} architecture consisting of multiple sensors and actuators (field instrumentation control layer), four \glspl{PLC} (process control layer), and a \gls{SCADA} workstation, including an \gls{HMI} and data historian (supervisory control layer).
Communication is conducted via the MODBUS TCP/IP protocol.
An additional Kali Linux machine is included for launching cyber attacks. 
The process consists of four stages, each of which is controlled by one of the four \glspl{PLC}. 

The dataset describes the normal operation of the system as well as several types of cyber attacks and physical faults against different components and communication links.
Among the cyber attacks are eight different \gls{MiTM} attacks, five \gls{DoS}, and seven scanning attacks. 
Moreover, three different water leaks and six sensor and pump breakdowns are included as physical events, which partly appear simultaneously and are interchangeably called attacks or faults by the authors.
In the present work, they will be referred to as physical faults. 
The dataset consists of two sub-datasets, namely a cyber network and physical process dataset.
While the physical dataset has a constant one-second resolution, the network dataset on average has 2633 observations per second. 
The raw features are listed in Table \ref{tab:features}.
For a more detailed explanation of the dataset the reader is referred to \cite{faramondi2021hardware}.
\begin{table}[h]
\caption{Raw cyber network and physical process features.}
\begin{center}
\begin{tabular}{c l c l}
\hline
\textbf{No.}&\textbf{Physical features}&\textbf{No.}&\textbf{Network features}\\
\hline
1 & Timestamp & 1 & Timestamp \\

\multirow{2}{*}{2-9} & \multirow{2}{*}{\shortstack[l]{Pressure sensor\\ value  of tank 1-8}} & 2-3 & IP address (src. \& dst.)$^{\mathrm{a}}$\\

 &  & 4-5 & MAC address (src. \& dst.)\\

10-15 & \begin{tabular}{@{}l@{}}State of pump 1-6\end{tabular} & 6-7 & Port (src. \& dst.)\\

\multirow{2}{*}{16-19} & \multirow{2}{*}{\shortstack[l]{Flow sensor value \\ of flow sensor 1-4 }} & 8 & Protocol\\

 & & 9 & TCP flags \\

20-41 & State of valve 1-22 & 10 & Payload size \\

 & & 11 & MODBUS function code \\

 & & 12 & MODBUS response value \\
 
 & & 13-14 & No. of packets$^{\mathrm{b}}$ (src. \& dst.) \\
\hline
\multicolumn{4}{l}{$^{\mathrm{a}}$Src. and dst. refer to source and destination, respectively.} \\
\multicolumn{4}{l}{$^{\mathrm{b}}$Refers to packets of the same device during the last two seconds.}
\end{tabular}
\label{tab:features}
\end{center}
\end{table}

\subsection{Transferability to control systems in the power sector} \label{sec:transferability}
The considered \gls{ICS} constitutes a generic test bed with characteristics potentially differing from its real-world counterparts.
To shed light on the transferability of the results of this study, a contextualization of the investigated \gls{ICS} is required.
Prominent representatives of real-world \glspl{ICS} are control systems in the power sector, such as distribution system operator's \gls{SCADA} or \glspl{SAS}, upon which the following comparison is based.
An overview is given in Table \ref{tab:transferability}.

An important commonality is the \gls{SCADA} architecture and its components.
Thus, the investigated attack scenarios, targeting various \gls{SCADA} components, provide realistic scenarios for both system types. 
Another similarity is the well-defined and steady configuration of the physical process and cyber network.
Moreover, both systems show continuous and repetitive patterns of the physical process either due to process cycles (generic \gls{ICS}) or seasonality (power systems).
Such deterministic configurations and patterns allow to model both systems with a set of physical and network features. 

Differences are mainly related to the physical process. 
In contrast to the investigated \gls{ICS}, power systems are subject to external influences such as weather and customer behavior, increasing process volatility.  
Moreover, compared to a distribution system operator's \gls{SCADA} system, the number of physical and network components is relatively small.  
Thus, a central difference is system complexity due to higher volatility and number of components in power systems. 

To conclude, the present generic \gls{ICS} can be considered a valuable test case for smaller control systems in the power sector such as \glspl{SAS}. 
However, due to a lack of system complexity, investigation of intrusion detection for larger \gls{SCADA} systems will require more extensive test beds. 

\setlength{\tabcolsep}{2.7pt}
\begin{table}[h]
\caption{Comparison of the investigated \gls{ICS} to control systems in the power sector.}
\begin{center}
\begin{tabular}{c l c l}
\hline
\textbf{No.}&\textbf{Similarities}& \textbf{No.} & \textbf{Differences} \\
\hline
1 & \begin{tabular}{@{}l@{}} \gls{SCADA} architecture \& components \end{tabular} & 1 &  \begin{tabular}{@{}l@{}}  Volatility \& trend pattern \end{tabular}\\

 2 & \begin{tabular}{@{}l@{}}Attack types \& target components\end{tabular} &  2 &   \begin{tabular}{@{}l@{}} Physical fault types \end{tabular}  \\

 3 & \begin{tabular}{@{}l@{}} Steady network \& process configu- \\ ration (e.g., IP addresses, protocols \\ \& number of physical devices) \end{tabular} & 3 &  \begin{tabular}{@{}l@{}} External impacts (e.g., \\ weather \& customer  \\  behavior) \end{tabular}  \\

4 & \begin{tabular}{@{}l@{}} Continuous, repetitive and thus deter-  \\ministic network \& process patterns \end{tabular}  & 4 & \begin{tabular}{@{}l@{}}  Type of communication \\ protocol   \end{tabular}  \\

5 & \begin{tabular}{@{}l@{}} Continuous, discrete \& categorical \\ features  \end{tabular}  & 5 &  \begin{tabular}{@{}l@{}}Number of physical \& \\ network components\end{tabular} \\

\hline
\end{tabular}
\label{tab:transferability}
\end{center}
\end{table}

%% file: 03_Methodology.tex
\section{Methodology} \label{sec:methodology}
This section first describes the preparation of the dataset. 
Thereafter, the data pipelines applied to the intrusion detection and classification problem are introduced. 

\subsection{Data preparation}
\begin{table*}[t]
\caption{Extracted network traffic and physical process features.}
\begin{center}
\begin{tabular}{c l l l}
\hline
\textbf{Feature no.}&\textbf{Extracted network features} & \textbf{Description} & \textbf{Underlying raw features$^{\mathrm{a}}$} \\
\hline
1 & Number of data transfers & Count of data packets transferred during the last second. & Raw network traffic data index \\
2-3 & \begin{tabular}{@{}l@{}} MAC/IP mismatch occurrence \end{tabular}& \begin{tabular}{@{}l@{}}Mismatch indication between the IP and MAC address \\ of at least one network device within the last second. \end{tabular}& \begin{tabular}{@{}l@{}} IP and MAC address\end{tabular} \\
4-14 & \begin{tabular}{@{}l@{}} Abnormal instance occurrence \end{tabular} & \begin{tabular}{@{}l@{}} Indication of an abnormal instance occurrence within the \\ respective raw network feature during the last second.\end{tabular} & \begin{tabular}{@{}l@{}} All raw network features except timestamp\\ and number of packets \end{tabular}\\
15-25 & \begin{tabular}{@{}l@{}} Number of abnormal instance \\ occurrences  \end{tabular} & \begin{tabular}{@{}l@{}}  Count of occurrences of abnormal instances within a sp-\\ecific raw network feature during the last second. \end{tabular} & \begin{tabular}{@{}l@{}} All raw network features except timestamp\\ and number of packets \end{tabular}\\
26-36 & \begin{tabular}{@{}l@{}} Number of normal instance \\occurrences \end{tabular} & \begin{tabular}{@{}l@{}}  Count of occurrences of normal instances within a speci-\\fic raw network feature during the last second. \end{tabular} & \begin{tabular}{@{}l@{}} All raw network features except timestamp\\ and number of packets \end{tabular}\\
37-106 & \begin{tabular}{@{}l@{}} Number of occurrences for\\ each instance \end{tabular}  & \begin{tabular}{@{}l@{}}Individual occurrences count for all instances of the res-\\pective raw network feature during the last second. \end{tabular}& \begin{tabular}{@{}l@{}} All raw network features except timestamp,\\ port and number of packets \end{tabular} \\
107-117 & \begin{tabular}{@{}l@{}} Number of different instances \end{tabular} &  \begin{tabular}{@{}l@{}} Count of distinct instances of a specific raw network fea-\\ture within the last second. \end{tabular} & \begin{tabular}{@{}l@{}} All raw network features except timestamp\\ and number of packets \end{tabular}\\
118-128 & \begin{tabular}{@{}l@{}} Number of NaN occurrences  \end{tabular} &  \begin{tabular}{@{}l@{}} Count of NaN occurrences within a raw network feature\\ during the last second.  \end{tabular} &  \begin{tabular}{@{}l@{}} All raw network features except timestamp\\ and number of packets \end{tabular} \\
129-131 & \begin{tabular}{@{}l@{}} Mean value \end{tabular}  & \begin{tabular}{@{}l@{}} Mean value of the respective raw network feature during\\ the last second. \end{tabular} & \begin{tabular}{@{}l@{}} Payload size and number of packets \end{tabular}\\
132-177 & \begin{tabular}{@{}l@{}} Number of different class-\\specific instances$^{\mathrm{b}}$ \end{tabular} &  \begin{tabular}{@{}l@{}} Count of distinct event class-specific instances of the res-\\spective raw network feature during the last second. \end{tabular} & \begin{tabular}{@{}l@{}} All raw network features except timestamp,\\ IP address and MAC address \end{tabular}\\
\hline
\textbf{Feature no.}&\textbf{Extracted physical features} & \textbf{Description} & \textbf{Underlying raw features} \\
\hline
178-185 & Pressure value of tank 1-8  & Raw pressure value of the respective tank. &  Pressure sensor value of tank 1-8 \\
186-191 & State of pump 1-6  & Raw state of the respective pump. &  State of pump 1-6  \\
192-195 &  \begin{tabular}{@{}l@{}} Value of flow sensor 1-4 \end{tabular} & Raw value of the respective flow sensor. &  Value of flow sensor 1-4  \\
196-217 & State of valve 1-22  & Raw state of the respective valve. & State of valve 1-22 \\
218 &  \begin{tabular}{@{}l@{}} Normal progress of a process \\ cycle \end{tabular} &  \begin{tabular}{@{}l@{}}Normal state of the current process cycle, defined on the\\ range between zero and one. \end{tabular}&  Pressure value of tank 1 \\
219 &  \begin{tabular}{@{}l@{}}Sine transformed normal pro-\\gress of a process cycle \end{tabular} & \begin{tabular}{@{}l@{}}Sine transformation of the normal progress of a process\\ cycle.\end{tabular} & Pressure value of tank 1 \\
220 &  \begin{tabular}{@{}l@{}}Cosine transformed normal\\ progress of a process cycle \end{tabular} & \begin{tabular}{@{}l@{}}Cosine transformation of the normal progress of a process\\ cycle.\end{tabular} & Pressure value of tank 1 \\
\hline
\multicolumn{4}{l}{$^{\mathrm{a}}$Source and destination considered in case of IP address, MAC address, port and number of packets.} \\
\multicolumn{4}{l}{$^{\mathrm{b}}$Event class-specific instances are defined only based on the training dataset.} \\
\end{tabular}
\label{tab:extracted_features}
\end{center}
\end{table*}

\subsubsection{Data partitioning} \label{sec:partitioning}
A good practice to test performance and generalization of a fully specified \gls{ML} model is the application to a holdout test dataset which stems from the same target distribution as the training set but was not previously seen \cite{friedman2001elements}. 
The present dataset describes attack or fault events in time series format, where a specific event corresponds to a sequence of observations.
As a specific event can only occur once, its entire sequence must either be placed in the training or test dataset.
Placing observations from a single event in both the training and test set will assume information from future events during model training.
Thus, data shuffling or \gls{CV}-based performance evaluation will result in an overly optimistic model performance assessment. 

In this work, the time series format of the investigated dataset is considered to perform a fair evaluation of the model performance.  
The entire sequences of the last two events of each attack or fault\footnote{Since there are only three water leak events in the dataset, some of which also occur simultaneously with sensor or pump breakdowns, all events of physical faults are combined into a \textit{physical fault} event class.} class are reserved for the test dataset and not considered during model training.
As a result, the training set consists of the first \SI{80}{\percent} of all normal operation observations, \SI{73.07}{\percent} of the \gls{DoS} observations, \SI{80.91}{\percent} of the \gls{MiTM} observations, \SI{77.75}{\percent} of the physical fault observations, and \SI{71.42}{\percent} of the scanning observations.
In this way, the risk for an overly optimistic performance assessment through data leakage is minimized, while a typical $75$/$25$ ratio between the training and test set can be maintained. 
Thus, future works examining the present dataset are encouraged to use the same partitioning in order to improve validity and comparison of results. 
\subsubsection{Feature extraction and fusion} \label{sec:feature_extraction}
Cyber-physical intrusion detection and classification simultaneously processes physical process and network traffic data. 
As these originate from different domains, they usually exhibit unequal characteristics such as observation rates and noise levels.
Thus, feature extraction from raw data typically requires different approaches for these two data sources.

In this work, the extraction of features from raw network traffic is realized by several sample statistics.
An overview of the considered statistics and resulting set of network features is given in Table \ref{tab:extracted_features}.
The selected statistics evaluate the traffic for each second.
The objective is to retain existing and extract additional information compared to consideration of individual data packets, while network and physical process features are aligned and the number of model executions reduced.
As discussed in Subsection \ref{sec:transferability}, \glspl{ICS} usually exhibit well-defined and static network configurations, which include, for example, fixed sets of IP addresses or ports.
This characteristic can be exploited by statistics which indicate occurrence of instances or instance combinations not present during normal operation.
Related features include MAC/IP address mismatch occurrence (Feature no. 2-3) and abnormal instance occurrence (Feature no. 4-14).
The set of normal instances of a raw network feature is extracted from the normal operation observations of the training dataset (see Subsection \ref{sec:partitioning}).
To retain the information detail of individual data packets, an abnormal instance (combination) occurrence is already indicated if a single packet is affected during the considered second.
Additional information is extracted by contextualizing packets within each second based on several counts and mean values (see Table \ref{tab:extracted_features}).
If, for example, only normal packets are received during a \gls{DoS} attack, but at an unusual rate, it can only be detected based on the additional information provided by the context of multiple packets. 

As discussed in Subsection \ref{sec:transferability}, the volatility, and hence noise level, of raw physical process features is comparatively small in the present case.
For that reason, no processing, such as data smoothing, is required and raw features can directly be used (see Table \ref{tab:extracted_features}).
In addition, the normal progress of a process cycle is extracted with the associated feature vector $P = \{p_{1},p_{2},...,p_{N} \ | \ p_{i} \in \mathbb{R} \forall i\}$ of length $N$. 
Values of $P$ are defined on the range $p_i \in [0,d]$, where $d$ corresponds to the usual duration of a process cycle, which is derived from pressure sensor values of tank 1 within the training dataset.
While $P$ can represent an expected progress of, for example, $10$ percentage points between $p=0.85d$ and $p=0.95d$, the same progress from $p=0.95d$ to $p=0.05d$ in the next process cycle is not described properly due to the jump discontinuity. 
To eliminate the discontinuity, and hence account for the cyclical nature of the process, the additional feature vectors $P_{\text{sin}}$ and $P_{\text{cos}}$ are extracted by applying sine and cosine transformation \cite{Chakraborty} on $P$ according to
\begin{equation}
      p_{\text{sin},i} = \sin\left(\frac{2\pi p_{i}}{d}\right),  \ \text{and} \  p_{\text{cos},i} = \cos\left(\frac{2\pi p_{i}}{d}\right),
\end{equation}
$\forall i \in [1,N]$. 
Note that cosine transformation is required as the sine function alone is not bijective, which would lead to ambiguity in the process cycle progress.

In total, 220 features are extracted (see Table \ref{tab:extracted_features}), some of which exhibit constant values and thus are non-informative.
After removing the non-informative features, the final dataset comprises 161 features and 9185 observations.

\subsection{Intrusion detection and classification data pipelines} \label{sec:pipelines}
\input{./Figures/Pipeline.tikz}

Supervised detection and classification of attacks and faults constitutes a highly imbalanced multi-class classification problem. 
To improve the classification performance, several up- and downstream data transformation steps are considered.
Typical steps comprise scaling, dimensionality reduction, undersampling and oversampling \cite{Fernndez}.
Together with classification and prediction filtering, these define the intrusion detection and classification data pipeline considered in this work (see Fig. \ref{fig:pipeline}).
Note that prediction filtering stems from result evaluation in Section \ref{sec:evaluation} and is not considered during model selection.
As depicted in Fig. \ref{fig:pipeline}, the data pipeline maps observations of a cyber-physical dataset to the considered event classes.
For each of the transformation steps, several candidate methods are considered, which represent the most widely used techniques.
Scaling candidates include feature standardization by removing the mean and scaling to unit variance, normalization to values between zero and one, and scaling to the maximum absolute value. 
\Gls{PCA} with Bayesian selection of the number of principle components is considered for dimensionality reduction \cite{minka2000automatic}. 
Methods for undersampling include \gls{IHT} and removal of Tomek Links, while \gls{SMOTE} and Borderline \gls{SMOTE} constitute the oversampling methods. 
Note that undersampling and oversampling is only applied on training data.
The classification models include a \gls{RF}, \gls{KNN}, a \gls{SVM} and an \gls{ANN}. 
Prediction filtering is realized by a moving majority filter which outputs the most frequent label of the past six predictions. 
As scanning attacks do not necessarily appear in sequences, they are excluded from the filtering process.
\setlength{\tabcolsep}{2.4pt}
\begin{table}[!b]
\caption{Method and hyperparameter selection results.}
\begin{center}
\begin{tabular}{l | l l l l}

               &  \multicolumn{4}{c}{\textbf{Cyber-physical intrusion detection \& classification pipelines}}  \\
\hline
\textbf{Step}&\textbf{RF}&\textbf{KNN}&\textbf{SVM}&\textbf{ANN}\\
\hline
\textbf{Scaling} &  Standard. & Standard. & Max. val. sc. &  Max. val. sc.\\
\textbf{Dim.\,red.}  & None & PCA & None & PCA\\
\textbf{Unders.} & None & None & None & None\\
\textbf{Overs.} & SMOTE & None & Bor. SMOTE & Bor. SMOTE\\
\textbf{Classif.} & \begin{tabular}{@{}l@{}} \textit{n}$_{\text{estimators}}$: \\ 100, \\ \textit{n}$_{\text{max-features}}$: \\ 17 \end{tabular}

& \begin{tabular}{@{}l@{}} \textit{n}$_{\text{neighbors}}$: 5, \\ \textit{Dist. func.}: \\ manhattan, \\ \textit{Weight func.}: \\ distance  \end{tabular}

&  \begin{tabular}{@{}l@{}} \textit{Kernel}: radial-\\basis function, \\ \textit{Penalty para.}: \\ 10000, \textit{Kernel} \\ \textit{coeff.}: 0.0175 \end{tabular} 

& \begin{tabular}{@{}l@{}} \textit{n}$_{\text{hid.-layers}}$: 2, \textit{n}$_{\text{units}}$:\\ 150,  \textit{Act. func.}: Re- \\Lu, \textit{Dropout rate}: \\ 0.5,  \textit{n}$_{\text{epochs}}$:  500, \\ \textit{Batch size}: 512 \end{tabular}\\

               &  \multicolumn{4}{c}{\textbf{Network intrusion detection \& classification pipelines}}  \\
\hline
\textbf{Step}&\textbf{RF}&\textbf{KNN}&\textbf{SVM}&\textbf{ANN}\\
\hline
\textbf{Scaling} &  Max.\,val.\,sc. & Standard. & Max. val. sc. &  Max. val. sc.\\
\textbf{Dim.\,red.}  & None & PCA & None & PCA\\
\textbf{Unders.} & IHT & Tomek Links & None & None\\
\textbf{Overs.} & None & None & None & Bor. SMOTE\\
\textbf{Classif.} & \begin{tabular}{@{}l@{}} \textit{n}$_{\text{estimators}}$: \\ 100, \\ \textit{n}$_{\text{max-features}}$: \\ 17 \end{tabular}

& \begin{tabular}{@{}l@{}} \textit{n}$_{\text{neighbors}}$: 5, \\ \textit{Dist. func.}: \\ manhattan, \\ \textit{Weight func.}: \\ uniform  \end{tabular}

&  \begin{tabular}{@{}l@{}} \textit{Kernel}: radial-\\basis function, \\ \textit{Penalty para.}: \\ 10000, \textit{Kernel} \\ \textit{coeff.}: 0.0175 \end{tabular} 

& \begin{tabular}{@{}l@{}} \textit{n}$_{\text{hid.-layers}}$: 2, \textit{n}$_{\text{units}}$:\\ 100,  \textit{Act. func.}: Re- \\Lu, \textit{Dropout rate}: \\ 0.5,  \textit{n}$_{\text{epochs}}$:  500, \\ \textit{Batch size}: 256 \end{tabular}\\
\hline
\end{tabular}
\label{tab:model_selection}
\end{center}
\end{table}
All data transformation steps, except for classification, can also be bypassed. 
For some classifiers specific transformation steps, such as bypassing scaling for \gls{SVM}, are excluded due to numerical issues.


The pipeline and most of the embedded data transformation methods and classification models are implemented in Python using the \textit{scikit-learn} library \cite{sklearndocumentation}.
An exception is the \gls{ANN} which is implemented using the deep learning library \textit{Keras} \cite{chollet2015keras}. 
Due to the multitude of models and techniques considered in this work, theoretical descriptions are omitted for brevity. 
Thus, for detailed backgrounds, the reader is referred to the respective library documentation as well as \cite{friedman2001elements} and \cite{Fernndez}.
The selection of transformation methods and hyperparameters is conducted based on the training set (see Section \ref{sec:partitioning}) applying a shuffled and stratified $5$-fold \gls{CV} grid search. 
Although shuffling introduces data leakage during model selection, it is required to ensure observations of each attack type in all folds.
While this may result in selection of non-optimal hyperparameters, the evaluation of the selected models is unaffected.
The method and hyperparameter selection is summarized in Table \ref{tab:model_selection}, where selected pipelines are referred to as the respective classifier. 
For hyperparameters not defined in Table \ref{tab:model_selection}, the library's default values are used. 
Before being applied to test data, the selected detection and classification pipelines are retrained on the full training set.

%% file: Figures/Pipeline.tikz
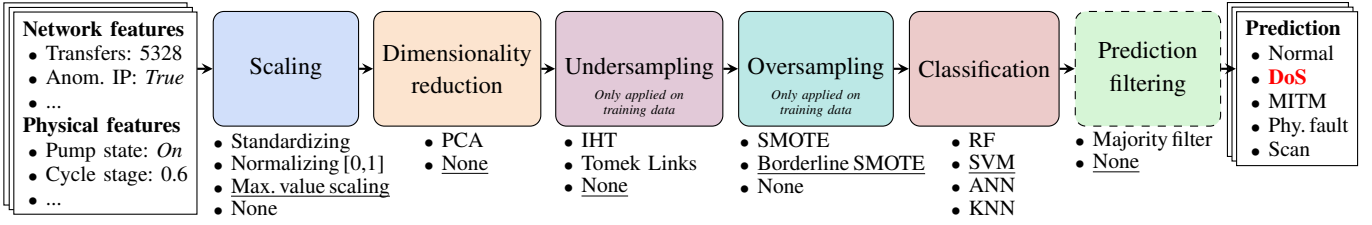
\begin{figure*}[t]
\centering
\resizebox{1\textwidth}{!}{
\begin{tikzpicture} 


\node[rectangle,draw, fill=white] at (-2.76,-0.46) {
    {\footnotesize\begin{varwidth}{\linewidth}
    \textbf{Network features}
    \begin{itemize}[leftmargin=0.3cm, labelsep=0.1cm, topsep=0cm]
        \item Transfers: 5328
        \item Anom. IP: \textit{True}
        \item ...
    \end{itemize}        
    \textbf{Physical features}
    \begin{itemize}[leftmargin=0.3cm, labelsep=0.1cm, topsep=0cm]
        \item Pump state: \textit{On}
        \item Cycle stage: 0.6
        \item ...
    \end{itemize}
    \end{varwidth}}
};

\node[rectangle,draw, fill=white] at (-2.69,-0.52) {
    {\footnotesize\begin{varwidth}{\linewidth}
    \textbf{Network features}
    \begin{itemize}[leftmargin=0.3cm, labelsep=0.1cm, topsep=0cm]
        \item Transfers: 5328
        \item Anom. IP: \textit{True}
        \item ...
    \end{itemize}        
    \textbf{Physical features}
    \begin{itemize}[leftmargin=0.3cm, labelsep=0.1cm, topsep=0cm]
        \item Pump state: \textit{On}
        \item Cycle stage: 0.6
        \item ...
    \end{itemize}
    \end{varwidth}}
};

\node[rectangle,draw, fill=white] at (-2.63,-0.58) {
    {\footnotesize\begin{varwidth}{\linewidth}
    \textbf{Network features}
    \begin{itemize}[leftmargin=0.3cm, labelsep=0.1cm, topsep=0cm]
        \item Transfers: 5328
        \item Anom. IP: \textit{True}
        \item ...
    \end{itemize}        
    \textbf{Physical features}
    \begin{itemize}[leftmargin=0.3cm, labelsep=0.1cm, topsep=0cm]
        \item Pump state: \textit{On}
        \item Cycle stage: 0.6
        \item ...
    \end{itemize}
    \end{varwidth}}
};

\node[rectangle,draw, fill=white] at (12.83,-0.14) {
    {\footnotesize\begin{varwidth}{\linewidth}
    \textbf{Prediction}
    \begin{itemize}[leftmargin=0.3cm, labelsep=0.1cm, topsep=0cm]
        \item Normal
        \item \gls{DoS}
        \item \gls{MiTM}
        \item Phy.\,fault
        \item Scan
    \end{itemize}        
    \end{varwidth}}
};

\node[rectangle,draw, fill=white] at (12.89,-0.20) {
    {\footnotesize\begin{varwidth}{\linewidth}
    \textbf{Prediction}
    \begin{itemize}[leftmargin=0.3cm, labelsep=0.1cm, topsep=0cm]
        \item Normal
        \item \gls{DoS}
        \item \gls{MiTM}
        \item Phy.\,fault
        \item Scan
    \end{itemize}        
    \end{varwidth}}
};

\node[rectangle,draw, fill=white] at (12.95,-0.26) {
    {\footnotesize\begin{varwidth}{\linewidth}
    \textbf{Prediction}
    \begin{itemize}[leftmargin=0.3cm, labelsep=0.1cm, topsep=0cm]
        \item Normal
        \item \textbf{\red{\gls{DoS}}}
        \item \gls{MiTM}
        \item Phy.\,fault
        \item Scan
    \end{itemize}        
    \end{varwidth}}
};




\node (Scaling) [Scaling, xshift=-0.27cm, align=center] { \small Scaling};

\node (PCA) [PCA, right of=Scaling, xshift=1.24cm, align=center] { \small Dimensionality \\ \small reduction};

\node (Undersampling) [Undersampling, right of=PCA, xshift=1.38cm,align=center] { \small  Undersampling};

\node (Oversampling) [Oversampling, right of=Undersampling, xshift=1.31cm, align=center] { \small Oversampling};

\node (Classification) [Classification, right of=Oversampling, xshift=1.2cm,align=center] { \small  Classification};

\node (Filtering) [Filtering, right of=Classification, xshift=1.14cm,align=center] { \small Prediction \\ \small filtering};

\draw [arrow] (-1.43,0) -- (Scaling);
\draw [arrow] (Scaling) -- (PCA);
\draw [arrow] (PCA) -- (Undersampling);
\draw [arrow] (Undersampling) -- (Oversampling);
\draw [arrow] (Oversampling) -- (Classification);
\draw [arrow] (Classification) -- (Filtering);
\draw [arrow] (Filtering) -- (12.15,0);


\node [below of=Scaling, yshift=-0.38cm, xshift=-0.03cm]{
    {\footnotesize\begin{varwidth}{\linewidth}
    \begin{itemize}[labelsep=0.1cm, topsep=0pt,itemsep=-0.02cm]
        \item Standardizing
        \item Normalizing\,[0,1]
        \item \underline{\smash{Max.\,value\,scaling}}
        \item None
    \end{itemize}
    \end{varwidth}}
};

\node  [below of=PCA, yshift=-0.11cm, xshift=-0.25cm] {
    {\footnotesize\begin{varwidth}{\linewidth}
    \begin{itemize}[labelsep=0.1cm, topsep=0pt,itemsep=-0.02cm]
        \item PCA
        \item \underline{None}
    \end{itemize}
    \end{varwidth}}
};

\node [below of=Undersampling, yshift=-0.26cm, xshift=-0.35cm] {
    {\footnotesize\begin{varwidth}{\linewidth}
    \begin{itemize}[labelsep=0.1cm, topsep=0pt,itemsep=-0.02cm]
        \item IHT
        \item Tomek Links
        \item \underline{None}
    \end{itemize}
    \end{varwidth}}
};

\node [below of=Oversampling, yshift=-0.23cm, xshift=-0.02cm] {
    {\footnotesize\begin{varwidth}{\linewidth}
    \begin{itemize}[labelsep=0.1cm, topsep=0pt,itemsep=-0.02cm]
        \item SMOTE
        \item \underline{Borderline\,SMOTE}
        \item None
    \end{itemize}
    \end{varwidth}}
};

\node  [below of=Classification, yshift=-0.37cm, xshift=-0.25cm] {
    {\footnotesize\begin{varwidth}{\linewidth}
    \begin{itemize}[labelsep=0.1cm, topsep=0pt,itemsep=-0.02cm]
        \item RF
        \item \underline{SVM}
        \item ANN
        \item KNN
    \end{itemize}
    \end{varwidth}}
};

\node  [below of=Filtering, yshift=-0.09cm, xshift=-0.29cm] {
    {\footnotesize\begin{varwidth}{\linewidth}
    \begin{itemize}[labelsep=0.1cm, topsep=0pt,itemsep=-0.02cm]
        \item Majority\,filter
        \item \underline{None}
    \end{itemize}
    \end{varwidth}}
};

\node[below of=Undersampling, yshift=0.65cm] {\tiny \textit{Only applied on}};
\node[below of=Undersampling, yshift=0.45cm] {\tiny \textit{training data}};

\node[below of=Oversampling, yshift=0.65cm] {\tiny \textit{Only applied on}};
\node[below of=Oversampling, yshift=0.45cm] {\tiny \textit{training data}};

\end{tikzpicture}}
\caption{Data pipeline for cyber-physical intrusion detection and multi-class classification, which maps observations of a dataset (left side) to event classes (right side). Underscores indicate an exemplary selection of methods for all data transformation steps.}
\label{fig:pipeline}
\end{figure*}

%% file: 04_Results.tex
\section{Performance evaluation} \label{sec:evaluation}
This section assesses the performance of cyber-physical intrusion detection and multi-class classification. 
Subsection \ref{sec:metrics} introduces the applied performance metrics. 
In Subsection \ref{sec:comparison} a comparison to network intrusion detection and classification is conducted, while Subsection \ref{sec:delay_misclass} evaluates detection delay and misclassifications.   

\subsection{Metrics} \label{sec:metrics}
To evaluate the class-wise detection and classification performance, this study considers the $F_{1}$ score according to
\begin{equation}
    F_{1,i} = \frac{TP_{i}}{TP_{i} + \frac{1}{2}(FP_{i} + FN_{i})},  \label{eq:F1}
\end{equation}
where $TP_{i}$, $FP_{i}$ and $FN_{i}$ are the number of true positives, false positives and false negatives of the $i$-th class, respectively. 
The overall performance is assessed based on a macro average of the class-wise $F_{1}$ scores, given as
\begin{equation}
    F^{m}_{1} = \frac{\sum^{N_{\text{classes}}}_{i=1}  F_{1,i}}{N_{\text{classes}}},  \label{eq:F1_macro}
\end{equation}
with $N_{\text{classes}}$ being the number of classes.
As seen from \eqref{eq:F1_macro}, the macro average $F^{m}_{1}$ treats all classes evenly, which is important given the high cost of missing observations of the less-populated attack classes. 
Average detection delay of class $i$ and average detection delay over all classes are given by
\begin{equation}
    \tau_{i}  =  \frac{\sum^{N_{\text{events},i}}_{j=1}(t_{\text{det},j,i} - t_{\text{start},j,i})}{N_{\text{events},i}},  \ \text{and} \  \tau =  \frac{\sum^{N_{\text{classes}}}_{i=1} \tau_{i}}{N_{\text{classes}}}, \label{eq:detection_delay}
\end{equation}
where $N_{\text{events},i}$ is the number of events,  $t_{\text{start},j,i}$ the start time of the $j$-th event and  $t_{\text{det},j,i}$ the first-detection time of the $j$-th event of the $i$-th class, respectively.

\subsection{Comparison of network and cyber-physical intrusion detection and multi-class classification}  \label{sec:comparison}
In Table \ref{tab:F1_results} the $F_{1}$ scores of all detection and classification pipelines are listed. 
The highest scores are in bold, while the second best scores are underlined. For network intrusion detection, all models show a similar overall performance, despite the differences on a class level.
As expected, physical faults are barely detected with pure network data.
However, most models detect some physical fault observations, especially the \gls{ANN}.
It can be concluded that network data provide some information about the physical process, which for instance could result from altered payload sizes or higher NaN occurrences.
\setlength{\tabcolsep}{4.2pt}
\begin{table}[b] 
\caption{Class-wise and average $F_{1}$ scores for network and cyber-physical intrusion detection and classification.}
\begin{center}
\begin{tabular}{l | c c c c |c c c c c}

               &  \multicolumn{4}{c|}{Network features} &  \multicolumn{4}{c}{Cyber-physical features} \\
\hline 
\textbf{Event class} &\textbf{\gls{RF}}&\textbf{\gls{KNN}}&\textbf{\gls{SVM}}&\textbf{\gls{ANN}}&\textbf{\gls{RF}}&\textbf{\gls{KNN}}&\textbf{\gls{SVM}}&\textbf{\gls{ANN}}\\
\hline
Normal & $0.92$  & $0.93$ & $0.86$ &  $0.88$ & $0.91$  & $\underline{0.94}$ & $\textbf{0.95}$ &  $0.93$\\
\gls{DoS} &  $0.55$ & $0.47$ & $\underline{0.96}$ &  $0.47$ & $0.71$ & $0.49$ & $\textbf{1.00}$ &  $0.50$\\
\gls{MiTM} &  $0.87$ & $0.83$ & $0.42$ & $0.68$  &  $0.87$ & $\underline{0.88}$ & $0.81$ &  $\textbf{0.92}$\\
Phy. fault & $0.07$ & $0.04$ & $0.00$ & $0.14$  & $0.26$  & $0.06$ & $\textbf{0.62}$ &  $\underline{0.46}$\\
Scanning & $0.57$  & $0.67$ & $0.80$ & $0.80$  & $\textbf{1.00}$  & $0.80$ & $\textbf{1.00}$ &  $\textbf{1.00}$\\
\hline
Avg. ($F^{m}_{1}$) &  $0.60$ & $0.59$ & $0.61$ & $0.60$  &  $0.75$ & $0.63$ & $\textbf{0.88}$ &  $\underline{0.76}$
\end{tabular}
\label{tab:F1_results}
\end{center}
\end{table}

The use of the cyber-physical feature set improves class-wise and overall performance for all models, with detection of normal observations by the \gls{RF} being the only exception. 
This allows the conclusion that incorporating physical process data has the potential to improve supervised intrusion detection and classification in \glspl{ICS}.
Interestingly, also the classification of scanning attacks improves, although these do not affect the physical process.
In fact, the absence of physical impact can be informative in the case where observations of \textit{different} attack types exhibit \textit{similar} impact on network traffic. 
If only one of the attack types also impacts the physical process, classification of both will be improved by incorporating physical data, due to less confusion between these attacks. 
As a result, physical process features also improve detection and classification of purely network traffic-affecting attacks.

The number of scanning attack observations is small compared to the other attack types.
Nevertheless, most of the studied cyber-physical pipelines perfectly detect and classify scanning attacks, despite the very few training examples.
It can be inferred that the extracted features in Table \ref{tab:extracted_features} very well capture the distinctive characteristics of scanning attacks.

Although \glspl{ANN} can capture highly non-linear and complex relationships, they achieve only second best performance.
An explanation may be the insufficient number of observations.
However, as training data of cyber attacks usually is scarce, \glspl{ANN} might not be appropriate for the given problem. 
The highest overall performance is achieved by the \gls{SVM}, due to the superior exploitation of physical process information.  
Moreover, \glspl{SVM} are known to be accurate also in high dimensional spaces, which may be an advantage given the relatively large feature-to-observation ratio. 
The good performance of the \gls{SVM} under existence of physical faults demonstrates that integration of physical process data also allows simultaneous detection and classification of events of fundamentally different nature, paving the way for holistic cross-domain root cause analysis. 
Nevertheless, the comparatively low $F_{1}$ score for physical faults requires further improvements such as comprehensive physical feature extraction.

While the overall performance ($F^{m}_{1}$) on average improves by $15.5$ percentage points, the models show very different class-specific improvements.  
Although detection of physical faults clearly improves for most models, \gls{KNN} sets an exception. 
Moreover, only the \gls{SVM} and \gls{ANN} show strong improvements for \gls{MiTM} detection, while the \gls{RF} shows a comparatively good improvement for \gls{DoS} attacks.
This complementarity suggests further investigation of ensemble modeling, e.g., combination of a \gls{SVM} and \gls{ANN}. 

\subsection{Misclassification and detection delay evaluation of cyber-physical intrusion detection and multi-class classification} \label{sec:delay_misclass}
Misclassifications and detection delay are investigated on the \gls{SVM} considering cyber-physical features due to superior performance. 
The confusion matrix in Fig. \ref{fig:Confusion_matrix} reveals that the \gls{SVM} mainly confuses physical faults and \gls{MiTM} attacks with normal operation and vice versa. 
Moreover, it shows almost no confusion among the attack and fault classes.
\input{Figures/Confusion.tikz} \input{Figures/Time_series_plot.tikz}
To further evaluate the location of misclassifications, the true and predicted labels of an excerpt of the test dataset are depicted in Fig. \ref{fig:time_series_plot}(a) in time series format. 
It can be noticed that misclassifications do not primarily appear during transition between event classes and are rather distributed. 
However, some increased emergence can be noticed at the transition from physical fault to normal operation (15:55:56) and at the beginning and end of the \gls{MiTM} attack between 16:08:26 and 16:10:01. 
High misclassification densities exist around 16:00:55 and 16:06:20 during normal operation, which might result from unlabeled irregular process behavior or noise.
Finally, most misclassifications occur individually and not in sequences.
This finding suggests filtering of the classification output (prediction filtering), as indicated in Fig. \ref{fig:pipeline}. 
Fig. \ref{fig:time_series_plot}(b) depicts the predictions after applying the majority filter described in Subsection \ref{sec:pipelines}.
The implemented filter is simple and not the result of a purely training data-based model selection process.
Thus, results are of preliminary nature and further investigation is required.
From a comparison of Fig. \ref{fig:time_series_plot}(a) and (b), it is noticeable that individual misclassifications are filtered out and only sequences remain, which greatly reduces the number of false positives. 
A quantitative assessment of the additional prediction filtering step in terms of $F_{1}$ score and detection delay for the test set excerpt of Fig. \ref{fig:time_series_plot} is given in Table \ref{tab:delay_f1}.
Prediction filtering improves overall detection performance ($F^{m}_{1}$) by $3$ percentage points. 
While detection of physical faults and \gls{MiTM} attacks greatly improves, the performance of \gls{DoS} detection decreases.
This can be explained by the prediction filtering-induced false negatives and positives at the beginning and end of the initially perfectly classified \gls{DoS} events. 
From Table \ref{tab:delay_f1} it can also be seen that the unfiltered \gls{SVM} immediately detects all event classes except for physical faults. 
Prediction filtering increases detection delay, since several seconds of an event need to elapse to reach majority within the filter sequence. 
\setlength{\tabcolsep}{4.5pt}
\begin{table}[h] 
\caption{Comparison of the unfiltered and filtered \gls{SVM} based on F1 score and detection delay for the time series depicted in Fig. 3.}
\begin{center}
\begin{tabular}{l | c |  c c c c c | c} 

\textbf{Model} & \textbf{Metric} &\textbf{Normal}&\textbf{\gls{DoS}}&\textbf{\gls{MiTM}}&\textbf{P.\,fault}&\textbf{Scan}&\textbf{Average}\\ 
\hline
\multirow{2}{*}{\shortstack[l]{ \gls{SVM} \\ unfilt.}} & $F_1$\,[-] & $0.94$ & $1.00$ &  $0.85$ & $0.62$  & $1.00$ & $0.88$ \\ 
 & $\tau_i$\,[s]  & $-$ & $0.00$ &  $0.00$ & $3.00$  & $0.00$ & $0.75$ \\ 
\hline
\multirow{2}{*}{\shortstack[l]{ \gls{SVM} \\ filtered}} & $F_1$\,[-] & $0.96$ & $0.90$ &  $0.98$ & $0.72$ & $1.00$ & $0.91$ \\ 
& $\tau_i$\,[s] & $-$ & $3.00$ &  $3.00$ & $6.00$ & $0.00$ & $3.00$\\ 
\end{tabular}
\label{tab:delay_f1}
\end{center}
\end{table}

%% file: Figures/Confusion.tikz
\def\myConfMat{{
{1460,0,45,41,0},  
{0,42,0,0,0},  
{14,0,128,0,0},  
{48,0,1,74,0},  
{0,0,0,0,2},  
}}
\def\classNames{{"\scriptsize\smash{Normal}","\scriptsize\smash{DoS}","\scriptsize\smash{MITM}","\scriptsize\smash{P.\,fault}", "\scriptsize\smash{Scanning}"}} 

\def\positions{{0,0,0,0,0}}

\def\numClasses{5} 
\def\myScale{0.885} 
\begin{figure}[h]
\centering
\begin{tikzpicture}[
    scale = \myScale,
    font={\small}, 
    ]

\tikzset{vertical label/.style={rotate=90,anchor=east}}   
\tikzset{diagonal label/.style={rotate=45,anchor=north east}}

\foreach \y in {1,...,\numClasses} 
{
    \pgfmathparse{-\y + {0.55,0.38,0.51,0.51,0.64}[\y-1]}
    \node [anchor=east, rotate=90,] at (0.4,\pgfmathresult) {\pgfmathparse{\classNames[\y-1]}\pgfmathresult}; 
    
    \foreach \x in {1,...,\numClasses}  
    {
    \def\totSamples{0}
    \foreach \ll in {1,...,\numClasses}
    {
        \pgfmathparse{\myConfMat[\y-1][\ll-1]}   
        \xdef\totSamples{\totSamples+\pgfmathresult} 
    }
    \pgfmathparse{\totSamples} \xdef\totSamples{\pgfmathresult}  
    
    \begin{scope}[shift={(\x,-\y)}]
        \def\mVal{\myConfMat[\y-1][\x-1]} 
        \pgfmathtruncatemacro{\r}{\mVal}   %
        \pgfmathtruncatemacro{\p}{round(\r/\totSamples*100)}
        \coordinate (C) at (0,0);
        \ifthenelse{\p<50}{\def\txtcol{black}}{\def\txtcol{white}} 
        \node[
            draw,                 
            text=\txtcol,         
            align=center,         
            fill=bblue!\p,        
            minimum size=\myScale*10mm,    
            inner sep=0,          
            ] (C) {\r\\\scriptsize\p\kern 0.1em\%};     
        \ifthenelse{\y=\numClasses}{
        \node [] at ($(C)-(0,0.75)$) 
        {\pgfmathparse{\classNames[\x-1]}\pgfmathresult};}{}
    \end{scope}
    }
}
\coordinate (yaxis) at (-0.15,0.8-\numClasses/2);  
\coordinate (xaxis) at (0.48+\numClasses/2, -\numClasses-1); 
\node [vertical label] at (yaxis) {True event class};
\node []               at (xaxis) {Predicted event class};
\end{tikzpicture}
    \caption{Confusion matrix of the \gls{SVM} for the cyber-physical feature set.} \label{fig:Confusion_matrix}
\end{figure}

%% file: Figures/Time_series_plot.tikz
\begin{figure}[t]
\centering

    \begin{tikzpicture}
    \begin{axis}[
         width  = 0.495*\textwidth,
        height = 3.37cm,
        xlabel = {\small (a) Unfiltered},
        xlabel style={yshift=13.5pt},
      legend image post style={scale=0.5},
      xtick pos=left,
      yticklabels={\footnotesize Normal, \footnotesize DoS,  \footnotesize MITM, \footnotesize Phy.\,fault, \footnotesize Scanning},
        xticklabels={\footnotesize 15:55:40, \footnotesize 15:59:32,  \footnotesize 16:03:24, \footnotesize 16:07:16, \footnotesize 16:11:08},
       every major tick/.append style={major tick length=3pt, black},
       major y tick style = transparent,
        xmin=0,
        xmax=1056,
        x tick label style={color=white},
        xtick={63, 295, 527, 759, 991},
        ytick={0,1,2,3,4},
        ymajorgrids=true,
        legend pos=south west,
        legend cell align=left,
        legend style={nodes={scale=0.65, transform shape}, at={(0.741,0.57)},},
        legend image post style={line width =1pt}
]

       

         \addplot[
    color=black,
    mark=none,
    line width=0.3mm,
    ]
    coordinates {(0, 3) (1, 3) (2, 3) (3, 3) (4, 3) (5, 3) (6, 3) (7, 3) (8, 3) (9, 3) (10, 3) (11, 3) (12, 3) (13, 3) (14, 3) (15, 3) (16, 3) (17, 3) (18, 3) (19, 3) (20, 3) (21, 3) (22, 3) (23, 3) (24, 3) (25, 3) (26, 3) (27, 3) (28, 3) (29, 3) (30, 3) (31, 3) (32, 3) (33, 3) (34, 3) (35, 3) (36, 3) (37, 3) (38, 3) (39, 3) (40, 3) (41, 3) (42, 3) (43, 3) (44, 3) (45, 3) (46, 3) (47, 3) (48, 3) (49, 3) (50, 3) (51, 3) (52, 3) (53, 3) (54, 3) (55, 3) (56, 3) (57, 3) (58, 3) (59, 3) (60, 3) (61, 3) (62, 3) (63, 3) (64, 3) (65, 3) (66, 3) (67, 3) (68, 3) (69, 3) (70, 3) (71, 3) (72, 3) (73, 3) (74, 3) (75, 3) (76, 3) (77, 3) (78, 3) (79, 3) (80, 3) (81, 0) (82, 0) (83, 0) (84, 0) (85, 0) (86, 0) (87, 0) (88, 0) (89, 0) (90, 0) (91, 0) (92, 0) (93, 0) (94, 0) (95, 0) (96, 0) (97, 0) (98, 0) (99, 0) (100, 0) (101, 0) (102, 0) (103, 0) (104, 0) (105, 0) (106, 0) (107, 0) (108, 0) (109, 0) (110, 0) (111, 0) (112, 0) (113, 0) (114, 0) (115, 0) (116, 0) (117, 0) (118, 0) (119, 0) (120, 0) (121, 0) (122, 0) (123, 0) (124, 0) (125, 0) (126, 0) (127, 0) (128, 0) (129, 0) (130, 0) (131, 0) (132, 0) (133, 0) (134, 0) (135, 0) (136, 0) (137, 0) (138, 0) (139, 0) (140, 0) (141, 0) (142, 0) (143, 0) (144, 0) (145, 0) (146, 0) (147, 0) (148, 0) (149, 0) (150, 0) (151, 0) (152, 0) (153, 0) (154, 0) (155, 0) (156, 0) (157, 0) (158, 0) (159, 0) (160, 0) (161, 0) (162, 0) (163, 0) (164, 0) (165, 0) (166, 0) (167, 0) (168, 0) (169, 0) (170, 0) (171, 0) (172, 0) (173, 0) (174, 0) (175, 0) (176, 0) (177, 0) (178, 0) (179, 0) (180, 0) (181, 0) (182, 0) (183, 0) (184, 0) (185, 0) (186, 0) (187, 0) (188, 0) (189, 0) (190, 0) (191, 0) (192, 0) (193, 0) (194, 0) (195, 0) (196, 0) (197, 0) (198, 0) (199, 0) (200, 0) (201, 0) (202, 0) (203, 0) (204, 0) (205, 0) (206, 0) (207, 0) (208, 0) (209, 0) (210, 0) (211, 0) (212, 0) (213, 0) (214, 0) (215, 0) (216, 0) (217, 0) (218, 0) (219, 0) (220, 0) (221, 0) (222, 0) (223, 0) (224, 0) (225, 0) (226, 0) (227, 0) (228, 0) (229, 0) (230, 0) (231, 0) (232, 0) (233, 0) (234, 0) (235, 0) (236, 0) (237, 0) (238, 0) (239, 0) (240, 0) (241, 4) (242, 0) (243, 0) (244, 0) (245, 0) (246, 0) (247, 0) (248, 0) (249, 0) (250, 0) (251, 0) (252, 0) (253, 0) (254, 0) (255, 0) (256, 0) (257, 0) (258, 0) (259, 0) (260, 0) (261, 0) (262, 0) (263, 0) (264, 0) (265, 0) (266, 0) (267, 0) (268, 0) (269, 0) (270, 0) (271, 0) (272, 0) (273, 0) (274, 0) (275, 0) (276, 0) (277, 0) (278, 0) (279, 0) (280, 0) (281, 0) (282, 0) (283, 0) (284, 0) (285, 0) (286, 0) (287, 0) (288, 0) (289, 0) (290, 0) (291, 0) (292, 0) (293, 0) (294, 0) (295, 0) (296, 0) (297, 0) (298, 0) (299, 0) (300, 0) (301, 0) (302, 0) (303, 0) (304, 0) (305, 0) (306, 0) (307, 0) (308, 0) (309, 0) (310, 0) (311, 0) (312, 0) (313, 0) (314, 0) (315, 0) (316, 0) (317, 0) (318, 1) (319, 1) (320, 1) (321, 1) (322, 1) (323, 1) (324, 1) (325, 1) (326, 1) (327, 1) (328, 1) (329, 1) (330, 1) (331, 1) (332, 0) (333, 0) (334, 0) (335, 0) (336, 0) (337, 0) (338, 0) (339, 0) (340, 0) (341, 0) (342, 0) (343, 0) (344, 0) (345, 0) (346, 0) (347, 0) (348, 0) (349, 0) (350, 0) (351, 0) (352, 0) (353, 0) (354, 0) (355, 0) (356, 0) (357, 0) (358, 0) (359, 0) (360, 0) (361, 0) (362, 0) (363, 0) (364, 0) (365, 0) (366, 0) (367, 0) (368, 0) (369, 0) (370, 0) (371, 0) (372, 0) (373, 0) (374, 0) (375, 0) (376, 0) (377, 0) (378, 0) (379, 0) (380, 0) (381, 0) (382, 0) (383, 0) (384, 0) (385, 0) (386, 0) (387, 0) (388, 0) (389, 0) (390, 0) (391, 0) (392, 0) (393, 0) (394, 0) (395, 0) (396, 0) (397, 0) (398, 0) (399, 0) (400, 0) (401, 0) (402, 0) (403, 0) (404, 0) (405, 0) (406, 0) (407, 0) (408, 0) (409, 0) (410, 0) (411, 0) (412, 0) (413, 0) (414, 0) (415, 0) (416, 0) (417, 0) (418, 0) (419, 0) (420, 0) (421, 0) (422, 0) (423, 0) (424, 0) (425, 0) (426, 0) (427, 0) (428, 0) (429, 0) (430, 0) (431, 0) (432, 0) (433, 0) (434, 0) (435, 0) (436, 0) (437, 0) (438, 0) (439, 0) (440, 0) (441, 0) (442, 0) (443, 0) (444, 0) (445, 0) (446, 0) (447, 0) (448, 0) (449, 0) (450, 0) (451, 0) (452, 0) (453, 0) (454, 0) (455, 0) (456, 0) (457, 0) (458, 0) (459, 0) (460, 0) (461, 0) (462, 0) (463, 0) (464, 0) (465, 0) (466, 0) (467, 0) (468, 0) (469, 0) (470, 0) (471, 0) (472, 0) (473, 0) (474, 0) (475, 0) (476, 0) (477, 0) (478, 0) (479, 0) (480, 0) (481, 0) (482, 0) (483, 0) (484, 0) (485, 0) (486, 0) (487, 0) (488, 0) (489, 0) (490, 0) (491, 0) (492, 0) (493, 0) (494, 0) (495, 0) (496, 0) (497, 0) (498, 0) (499, 0) (500, 0) (501, 0) (502, 0) (503, 0) (504, 0) (505, 0) (506, 0) (507, 0) (508, 0) (509, 0) (510, 0) (511, 0) (512, 0) (513, 0) (514, 0) (515, 0) (516, 0) (517, 0) (518, 0) (519, 0) (520, 0) (521, 0) (522, 0) (523, 0) (524, 0) (525, 0) (526, 0) (527, 0) (528, 0) (529, 0) (530, 0) (531, 0) (532, 0) (533, 0) (534, 0) (535, 0) (536, 0) (537, 0) (538, 0) (539, 0) (540, 0) (541, 0) (542, 0) (543, 0) (544, 0) (545, 0) (546, 0) (547, 0) (548, 0) (549, 0) (550, 0) (551, 0) (552, 0) (553, 0) (554, 0) (555, 0) (556, 0) (557, 0) (558, 0) (559, 0) (560, 0) (561, 0) (562, 0) (563, 0) (564, 0) (565, 0) (566, 0) (567, 0) (568, 0) (569, 0) (570, 0) (571, 0) (572, 0) (573, 0) (574, 0) (575, 0) (576, 0) (577, 0) (578, 0) (579, 0) (580, 0) (581, 0) (582, 0) (583, 0) (584, 0) (585, 0) (586, 0) (587, 0) (588, 0) (589, 0) (590, 0) (591, 0) (592, 0) (593, 0) (594, 0) (595, 0) (596, 0) (597, 0) (598, 0) (599, 0) (600, 4) (601, 0) (602, 0) (603, 0) (604, 0) (605, 0) (606, 0) (607, 0) (608, 0) (609, 0) (610, 0) (611, 0) (612, 0) (613, 0) (614, 0) (615, 0) (616, 0) (617, 0) (618, 0) (619, 0) (620, 0) (621, 0) (622, 0) (623, 0) (624, 0) (625, 0) (626, 0) (627, 0) (628, 0) (629, 0) (630, 0) (631, 0) (632, 0) (633, 0) (634, 0) (635, 0) (636, 0) (637, 0) (638, 0) (639, 0) (640, 0) (641, 0) (642, 0) (643, 0) (644, 0) (645, 0) (646, 0) (647, 0) (648, 0) (649, 0) (650, 0) (651, 0) (652, 0) (653, 0) (654, 0) (655, 0) (656, 0) (657, 0) (658, 0) (659, 0) (660, 0) (661, 0) (662, 0) (663, 0) (664, 0) (665, 0) (666, 0) (667, 0) (668, 0) (669, 0) (670, 0) (671, 0) (672, 0) (673, 0) (674, 0) (675, 0) (676, 0) (677, 0) (678, 0) (679, 0) (680, 0) (681, 0) (682, 0) (683, 0) (684, 0) (685, 0) (686, 0) (687, 0) (688, 0) (689, 0) (690, 0) (691, 0) (692, 0) (693, 0) (694, 0) (695, 0) (696, 0) (697, 0) (698, 0) (699, 0) (700, 0) (701, 0) (702, 0) (703, 0) (704, 0) (705, 0) (706, 0) (707, 0) (708, 0) (709, 0) (710, 0) (711, 0) (712, 0) (713, 0) (714, 0) (715, 0) (716, 0) (717, 0) (718, 0) (719, 0) (720, 0) (721, 0) (722, 0) (723, 0) (724, 0) (725, 0) (726, 0) (727, 0) (728, 0) (729, 0) (730, 0) (731, 0) (732, 0) (733, 0) (734, 0) (735, 0) (736, 0) (737, 0) (738, 0) (739, 0) (740, 0) (741, 0) (742, 0) (743, 0) (744, 0) (745, 0) (746, 0) (747, 0) (748, 0) (749, 0) (750, 0) (751, 0) (752, 0) (753, 0) (754, 0) (755, 0) (756, 0) (757, 0) (758, 0) (759, 0) (760, 0) (761, 0) (762, 0) (763, 0) (764, 0) (765, 0) (766, 0) (767, 0) (768, 0) (769, 0) (770, 0) (771, 0) (772, 0) (773, 0) (774, 0) (775, 0) (776, 0) (777, 0) (778, 0) (779, 0) (780, 0) (781, 0) (782, 0) (783, 0) (784, 0) (785, 0) (786, 0) (787, 0) (788, 0) (789, 0) (790, 0) (791, 0) (792, 0) (793, 0) (794, 0) (795, 0) (796, 0) (797, 0) (798, 0) (799, 0) (800, 0) (801, 0) (802, 0) (803, 0) (804, 0) (805, 0) (806, 0) (807, 0) (808, 0) (809, 0) (810, 0) (811, 0) (812, 0) (813, 0) (814, 0) (815, 0) (816, 0) (817, 0) (818, 0) (819, 0) (820, 0) (821, 0) (822, 0) (823, 0) (824, 0) (825, 0) (826, 0) (827, 2) (828, 2) (829, 2) (830, 2) (831, 2) (832, 2) (833, 2) (834, 2) (835, 2) (836, 2) (837, 2) (838, 2) (839, 2) (840, 2) (841, 2) (842, 2) (843, 2) (844, 2) (845, 2) (846, 2) (847, 2) (848, 2) (849, 2) (850, 2) (851, 2) (852, 2) (853, 2) (854, 2) (855, 2) (856, 2) (857, 2) (858, 2) (859, 2) (860, 2) (861, 2) (862, 2) (863, 2) (864, 2) (865, 2) (866, 2) (867, 2) (868, 2) (869, 2) (870, 2) (871, 2) (872, 2) (873, 2) (874, 2) (875, 2) (876, 2) (877, 2) (878, 2) (879, 2) (880, 2) (881, 2) (882, 2) (883, 2) (884, 2) (885, 2) (886, 2) (887, 2) (888, 2) (889, 2) (890, 2) (891, 2) (892, 2) (893, 2) (894, 2) (895, 2) (896, 2) (897, 2) (898, 2) (899, 2) (900, 2) (901, 2) (902, 2) (903, 2) (904, 2) (905, 2) (906, 2) (907, 2) (908, 2) (909, 2) (910, 2) (911, 2) (912, 2) (913, 2) (914, 2) (915, 2) (916, 2) (917, 2) (918, 2) (919, 2) (920, 2) (921, 2) (922, 2) (923, 0) (924, 0) (925, 0) (926, 0) (927, 0) (928, 0) (929, 0) (930, 0) (931, 0) (932, 0) (933, 0) (934, 0) (935, 0) (936, 0) (937, 0) (938, 0) (939, 0) (940, 0) (941, 0) (942, 0) (943, 0) (944, 0) (945, 0) (946, 0) (947, 0) (948, 0) (949, 0) (950, 0) (951, 0) (952, 0) (953, 0) (954, 0) (955, 0) (956, 0) (957, 0) (958, 0) (959, 0) (960, 0) (961, 0) (962, 0) (963, 0) (964, 0) (965, 0) (966, 0) (967, 0) (968, 0) (969, 0) (970, 0) (971, 0) (972, 0) (973, 0) (974, 0) (975, 0) (976, 0) (977, 0) (978, 0) (979, 0) (980, 0) (981, 0) (982, 0) (983, 0) (984, 0) (985, 0) (986, 0) (987, 0) (988, 0) (989, 0) (990, 0) (991, 0) (992, 0) (993, 0) (994, 0) (995, 0) (996, 0) (997, 0) (998, 0) (999, 0) (1000, 0) (1001, 0) (1002, 0) (1003, 0) (1004, 0) (1005, 0) (1006, 0) (1007, 0) (1008, 0) (1009, 0) (1010, 0) (1011, 0) (1012, 0) (1013, 0) (1014, 0) (1015, 0) (1016, 0) (1017, 0) (1018, 0) (1019, 0) (1020, 0) (1021, 0) (1022, 0) (1023, 0) (1024, 0) (1025, 0) (1026, 0) (1027, 1) (1028, 1) (1029, 1) (1030, 1) (1031, 1) (1032, 1) (1033, 1) (1034, 1) (1035, 1) (1036, 1) (1037, 1) (1038, 1) (1039, 1) (1040, 1) (1041, 1) (1042, 1) (1043, 1) (1044, 1) (1045, 1) (1046, 1) (1047, 1) (1048, 1) (1049, 1) (1050, 1) (1051, 1) (1052, 1) (1053, 1) (1054, 1)};

    \addlegendentry{Ground truth}

    \addplot[
    color=rred,
    mark=none,
    line width=0.1mm,
    ]
    coordinates {(0, 0) (1, 0) (2, 0) (3, 3) (4, 3) (5, 3) (6, 3) (7, 3) (8, 3) (9, 0) (10, 0) (11, 3) (12, 3) (13, 3) (14, 3) (15, 3) (16, 3) (17, 3) (18, 3) (19, 3) (20, 3) (21, 0) (22, 3) (23, 3) (24, 2) (25, 3) (26, 3) (27, 0) (28, 3) (29, 3) (30, 3) (31, 3) (32, 0) (33, 0) (34, 3) (35, 3) (36, 3) (37, 3) (38, 3) (39, 3) (40, 3) (41, 3) (42, 3) (43, 3) (44, 3) (45, 3) (46, 3) (47, 3) (48, 3) (49, 3) (50, 3) (51, 3) (52, 3) (53, 3) (54, 3) (55, 3) (56, 0) (57, 3) (58, 3) (59, 0) (60, 0) (61, 3) (62, 0) (63, 0) (64, 3) (65, 3) (66, 3) (67, 3) (68, 3) (69, 3) (70, 0) (71, 0) (72, 0) (73, 0) (74, 0) (75, 0) (76, 0) (77, 0) (78, 0) (79, 0) (80, 0) (81, 0) (82, 0) (83, 0) (84, 0) (85, 0) (86, 0) (87, 0) (88, 0) (89, 0) (90, 0) (91, 0) (92, 0) (93, 0) (94, 0) (95, 0) (96, 0) (97, 0) (98, 0) (99, 0) (100, 0) (101, 0) (102, 0) (103, 0) (104, 0) (105, 0) (106, 0) (107, 0) (108, 0) (109, 0) (110, 0) (111, 0) (112, 0) (113, 0) (114, 0) (115, 0) (116, 0) (117, 0) (118, 0) (119, 0) (120, 0) (121, 0) (122, 0) (123, 0) (124, 0) (125, 0) (126, 0) (127, 0) (128, 0) (129, 0) (130, 0) (131, 0) (132, 0) (133, 0) (134, 0) (135, 0) (136, 0) (137, 0) (138, 0) (139, 0) (140, 0) (141, 0) (142, 0) (143, 0) (144, 0) (145, 0) (146, 0) (147, 0) (148, 0) (149, 0) (150, 0) (151, 0) (152, 0) (153, 0) (154, 0) (155, 0) (156, 0) (157, 0) (158, 0) (159, 0) (160, 0) (161, 0) (162, 0) (163, 0) (164, 0) (165, 0) (166, 0) (167, 0) (168, 0) (169, 0) (170, 0) (171, 0) (172, 0) (173, 0) (174, 0) (175, 0) (176, 0) (177, 0) (178, 0) (179, 0) (180, 0) (181, 0) (182, 0) (183, 0) (184, 0) (185, 0) (186, 0) (187, 0) (188, 2) (189, 0) (190, 0) (191, 0) (192, 2) (193, 0) (194, 0) (195, 0) (196, 0) (197, 2) (198, 0) (199, 2) (200, 0) (201, 0) (202, 0) (203, 0) (204, 0) (205, 0) (206, 0) (207, 0) (208, 0) (209, 2) (210, 0) (211, 0) (212, 0) (213, 0) (214, 0) (215, 0) (216, 0) (217, 0) (218, 0) (219, 0) (220, 0) (221, 0) (222, 0) (223, 0) (224, 0) (225, 0) (226, 0) (227, 0) (228, 0) (229, 0) (230, 0) (231, 0) (232, 0) (233, 0) (234, 0) (235, 0) (236, 0) (237, 0) (238, 0) (239, 0) (240, 0) (241, 4) (242, 0) (243, 0) (244, 0) (245, 0) (246, 0) (247, 0) (248, 0) (249, 0) (250, 0) (251, 0) (252, 0) (253, 2) (254, 0) (255, 0) (256, 0) (257, 0) (258, 0) (259, 0) (260, 0) (261, 0) (262, 0) (263, 0) (264, 0) (265, 0) (266, 0) (267, 0) (268, 0) (269, 0) (270, 0) (271, 0) (272, 0) (273, 0) (274, 0) (275, 0) (276, 0) (277, 0) (278, 0) (279, 0) (280, 0) (281, 0) (282, 0) (283, 0) (284, 0) (285, 0) (286, 3) (287, 0) (288, 0) (289, 0) (290, 0) (291, 0) (292, 0) (293, 0) (294, 0) (295, 0) (296, 0) (297, 0) (298, 0) (299, 0) (300, 0) (301, 0) (302, 0) (303, 0) (304, 0) (305, 0) (306, 0) (307, 0) (308, 0) (309, 0) (310, 0) (311, 0) (312, 0) (313, 0) (314, 0) (315, 0) (316, 0) (317, 0) (318, 1) (319, 1) (320, 1) (321, 1) (322, 1) (323, 1) (324, 1) (325, 1) (326, 1) (327, 1) (328, 1) (329, 1) (330, 1) (331, 1) (332, 0) (333, 0) (334, 0) (335, 0) (336, 0) (337, 0) (338, 0) (339, 0) (340, 0) (341, 0) (342, 0) (343, 0) (344, 0) (345, 0) (346, 0) (347, 0) (348, 0) (349, 0) (350, 0) (351, 0) (352, 0) (353, 3) (354, 3) (355, 0) (356, 0) (357, 0) (358, 0) (359, 3) (360, 0) (361, 0) (362, 3) (363, 0) (364, 0) (365, 0) (366, 0) (367, 0) (368, 0) (369, 0) (370, 0) (371, 0) (372, 3) (373, 0) (374, 0) (375, 0) (376, 0) (377, 0) (378, 0) (379, 0) (380, 0) (381, 3) (382, 0) (383, 3) (384, 3) (385, 3) (386, 3) (387, 3) (388, 3) (389, 3) (390, 3) (391, 3) (392, 3) (393, 3) (394, 3) (395, 3) (396, 3) (397, 3) (398, 3) (399, 3) (400, 3) (401, 3) (402, 3) (403, 0) (404, 0) (405, 0) (406, 0) (407, 0) (408, 0) (409, 0) (410, 0) (411, 0) (412, 0) (413, 0) (414, 0) (415, 0) (416, 0) (417, 0) (418, 0) (419, 0) (420, 0) (421, 0) (422, 0) (423, 0) (424, 0) (425, 0) (426, 0) (427, 0) (428, 0) (429, 0) (430, 0) (431, 0) (432, 0) (433, 0) (434, 0) (435, 0) (436, 0) (437, 0) (438, 0) (439, 0) (440, 0) (441, 0) (442, 0) (443, 0) (444, 0) (445, 0) (446, 0) (447, 0) (448, 0) (449, 0) (450, 0) (451, 0) (452, 0) (453, 0) (454, 0) (455, 0) (456, 0) (457, 0) (458, 0) (459, 0) (460, 0) (461, 0) (462, 0) (463, 0) (464, 0) (465, 0) (466, 0) (467, 0) (468, 0) (469, 0) (470, 0) (471, 0) (472, 0) (473, 0) (474, 0) (475, 0) (476, 0) (477, 0) (478, 0) (479, 0) (480, 0) (481, 0) (482, 0) (483, 0) (484, 0) (485, 0) (486, 0) (487, 0) (488, 0) (489, 0) (490, 0) (491, 0) (492, 0) (493, 0) (494, 0) (495, 0) (496, 0) (497, 0) (498, 0) (499, 0) (500, 0) (501, 0) (502, 0) (503, 2) (504, 0) (505, 0) (506, 0) (507, 0) (508, 0) (509, 0) (510, 0) (511, 0) (512, 0) (513, 0) (514, 0) (515, 0) (516, 0) (517, 0) (518, 0) (519, 0) (520, 0) (521, 0) (522, 0) (523, 0) (524, 0) (525, 0) (526, 0) (527, 0) (528, 0) (529, 0) (530, 0) (531, 0) (532, 0) (533, 0) (534, 0) (535, 0) (536, 0) (537, 0) (538, 0) (539, 0) (540, 0) (541, 0) (542, 0) (543, 0) (544, 0) (545, 0) (546, 2) (547, 0) (548, 0) (549, 0) (550, 0) (551, 0) (552, 0) (553, 0) (554, 0) (555, 0) (556, 0) (557, 0) (558, 0) (559, 0) (560, 0) (561, 2) (562, 0) (563, 0) (564, 0) (565, 0) (566, 0) (567, 2) (568, 0) (569, 2) (570, 0) (571, 0) (572, 0) (573, 0) (574, 0) (575, 0) (576, 0) (577, 2) (578, 0) (579, 0) (580, 0) (581, 0) (582, 0) (583, 0) (584, 0) (585, 0) (586, 0) (587, 0) (588, 0) (589, 0) (590, 0) (591, 0) (592, 0) (593, 0) (594, 0) (595, 0) (596, 0) (597, 0) (598, 0) (599, 0) (600, 4) (601, 0) (602, 0) (603, 0) (604, 0) (605, 0) (606, 0) (607, 0) (608, 0) (609, 0) (610, 0) (611, 0) (612, 0) (613, 0) (614, 0) (615, 0) (616, 0) (617, 0) (618, 0) (619, 0) (620, 0) (621, 0) (622, 0) (623, 0) (624, 0) (625, 0) (626, 0) (627, 0) (628, 0) (629, 0) (630, 0) (631, 0) (632, 0) (633, 0) (634, 0) (635, 0) (636, 0) (637, 0) (638, 0) (639, 0) (640, 0) (641, 0) (642, 0) (643, 0) (644, 0) (645, 0) (646, 0) (647, 0) (648, 0) (649, 0) (650, 0) (651, 0) (652, 0) (653, 0) (654, 0) (655, 0) (656, 0) (657, 0) (658, 0) (659, 3) (660, 3) (661, 0) (662, 0) (663, 3) (664, 0) (665, 0) (666, 0) (667, 0) (668, 3) (669, 0) (670, 0) (671, 0) (672, 0) (673, 3) (674, 0) (675, 0) (676, 0) (677, 0) (678, 0) (679, 0) (680, 0) (681, 3) (682, 3) (683, 3) (684, 3) (685, 3) (686, 3) (687, 3) (688, 0) (689, 0) (690, 0) (691, 3) (692, 0) (693, 0) (694, 3) (695, 0) (696, 0) (697, 0) (698, 0) (699, 0) (700, 0) (701, 0) (702, 0) (703, 2) (704, 0) (705, 0) (706, 2) (707, 0) (708, 0) (709, 2) (710, 0) (711, 0) (712, 0) (713, 0) (714, 2) (715, 0) (716, 0) (717, 0) (718, 0) (719, 2) (720, 0) (721, 0) (722, 2) (723, 2) (724, 0) (725, 0) (726, 0) (727, 0) (728, 0) (729, 0) (730, 0) (731, 0) (732, 2) (733, 0) (734, 0) (735, 0) (736, 0) (737, 0) (738, 0) (739, 0) (740, 2) (741, 2) (742, 0) (743, 0) (744, 0) (745, 0) (746, 2) (747, 2) (748, 0) (749, 0) (750, 0) (751, 0) (752, 0) (753, 0) (754, 0) (755, 0) (756, 0) (757, 0) (758, 0) (759, 0) (760, 0) (761, 0) (762, 0) (763, 0) (764, 0) (765, 0) (766, 0) (767, 0) (768, 0) (769, 0) (770, 0) (771, 0) (772, 0) (773, 0) (774, 0) (775, 0) (776, 0) (777, 0) (778, 0) (779, 0) (780, 0) (781, 0) (782, 0) (783, 0) (784, 0) (785, 0) (786, 0) (787, 0) (788, 0) (789, 0) (790, 0) (791, 0) (792, 0) (793, 0) (794, 0) (795, 0) (796, 0) (797, 0) (798, 0) (799, 0) (800, 0) (801, 0) (802, 0) (803, 0) (804, 0) (805, 0) (806, 0) (807, 0) (808, 0) (809, 0) (810, 0) (811, 0) (812, 0) (813, 0) (814, 0) (815, 0) (816, 0) (817, 0) (818, 0) (819, 0) (820, 0) (821, 0) (822, 0) (823, 0) (824, 0) (825, 0) (826, 0) (827, 2) (828, 2) (829, 2) (830, 2) (831, 2) (832, 2) (833, 2) (834, 2) (835, 0) (836, 2) (837, 2) (838, 2) (839, 0) (840, 2) (841, 2) (842, 2) (843, 2) (844, 2) (845, 2) (846, 2) (847, 2) (848, 2) (849, 2) (850, 2) (851, 2) (852, 2) (853, 2) (854, 2) (855, 2) (856, 2) (857, 2) (858, 2) (859, 2) (860, 2) (861, 2) (862, 2) (863, 0) (864, 2) (865, 2) (866, 2) (867, 2) (868, 2) (869, 2) (870, 2) (871, 2) (872, 2) (873, 2) (874, 2) (875, 2) (876, 2) (877, 2) (878, 2) (879, 2) (880, 2) (881, 2) (882, 2) (883, 2) (884, 2) (885, 2) (886, 2) (887, 0) (888, 2) (889, 2) (890, 2) (891, 2) (892, 2) (893, 2) (894, 2) (895, 2) (896, 2) (897, 2) (898, 2) (899, 2) (900, 2) (901, 2) (902, 2) (903, 2) (904, 2) (905, 2) (906, 2) (907, 2) (908, 2) (909, 2) (910, 2) (911, 2) (912, 2) (913, 2) (914, 2) (915, 2) (916, 0) (917, 2) (918, 2) (919, 2) (920, 0) (921, 0) (922, 2) (923, 0) (924, 0) (925, 0) (926, 0) (927, 0) (928, 0) (929, 0) (930, 0) (931, 0) (932, 0) (933, 0) (934, 0) (935, 0) (936, 0) (937, 0) (938, 0) (939, 0) (940, 0) (941, 0) (942, 0) (943, 0) (944, 0) (945, 0) (946, 0) (947, 0) (948, 0) (949, 0) (950, 0) (951, 0) (952, 0) (953, 0) (954, 0) (955, 0) (956, 0) (957, 0) (958, 0) (959, 0) (960, 0) (961, 0) (962, 0) (963, 0) (964, 0) (965, 0) (966, 0) (967, 0) (968, 0) (969, 0) (970, 0) (971, 0) (972, 0) (973, 0) (974, 0) (975, 0) (976, 0) (977, 0) (978, 0) (979, 0) (980, 0) (981, 0) (982, 0) (983, 0) (984, 0) (985, 0) (986, 0) (987, 0) (988, 0) (989, 0) (990, 0) (991, 0) (992, 0) (993, 0) (994, 0) (995, 0) (996, 0) (997, 0) (998, 0) (999, 0) (1000, 0) (1001, 0) (1002, 0) (1003, 0) (1004, 0) (1005, 0) (1006, 0) (1007, 0) (1008, 0) (1009, 0) (1010, 0) (1011, 0) (1012, 0) (1013, 0) (1014, 0) (1015, 0) (1016, 0) (1017, 0) (1018, 0) (1019, 0) (1020, 0) (1021, 0) (1022, 0) (1023, 0) (1024, 0) (1025, 0) (1026, 0) (1027, 1) (1028, 1) (1029, 1) (1030, 1) (1031, 1) (1032, 1) (1033, 1) (1034, 1) (1035, 1) (1036, 1) (1037, 1) (1038, 1) (1039, 1) (1040, 1) (1041, 1) (1042, 1) (1043, 1) (1044, 1) (1045, 1) (1046, 1) (1047, 1) (1048, 1) (1049, 1) (1050, 1) (1051, 1) (1052, 1) (1053, 1) (1054, 1)};
    
        \addlegendentry{Prediction}
        

\end{axis}

\end{tikzpicture}\vspace{-0.15cm}
\begin{tikzpicture}

    \begin{axis}[
         width  = 0.495*\textwidth,
        height = 3.37cm,
      xlabel = {\small (b) Filtered},
              xlabel style={yshift=5pt},
      legend image post style={scale=0.5},
      xtick pos=left,
      yticklabels={\footnotesize Normal, \footnotesize DoS,  \footnotesize MITM, \footnotesize Phy.\,fault, \footnotesize Scanning},
        xticklabels={\footnotesize 15:55:42, \footnotesize 15:59:34,  \footnotesize 16:03:26, \footnotesize 16:07:18, \footnotesize 16:11:10},
       every major tick/.append style={major tick length=3pt, black},
       major y tick style = transparent,
        xmin=0,
        xmax=1056,
        xtick={63, 295, 527, 759, 991},
        ytick={0,1,2,3,4},
        ymajorgrids=true,
        legend pos=south west,
        legend cell align=left,
        legend style={nodes={scale=0.6, transform shape}, at={(0.745,0.67)},},
        legend image post style={line width =1pt}
]

       

         \addplot[
    color=black,
    mark=none,
    line width=0.3mm,
    ]
    coordinates {(0, 3) (1, 3) (2, 3) (3, 3) (4, 3) (5, 3) (6, 3) (7, 3) (8, 3) (9, 3) (10, 3) (11, 3) (12, 3) (13, 3) (14, 3) (15, 3) (16, 3) (17, 3) (18, 3) (19, 3) (20, 3) (21, 3) (22, 3) (23, 3) (24, 3) (25, 3) (26, 3) (27, 3) (28, 3) (29, 3) (30, 3) (31, 3) (32, 3) (33, 3) (34, 3) (35, 3) (36, 3) (37, 3) (38, 3) (39, 3) (40, 3) (41, 3) (42, 3) (43, 3) (44, 3) (45, 3) (46, 3) (47, 3) (48, 3) (49, 3) (50, 3) (51, 3) (52, 3) (53, 3) (54, 3) (55, 3) (56, 3) (57, 3) (58, 3) (59, 3) (60, 3) (61, 3) (62, 3) (63, 3) (64, 3) (65, 3) (66, 3) (67, 3) (68, 3) (69, 3) (70, 3) (71, 3) (72, 3) (73, 3) (74, 3) (75, 3) (76, 3) (77, 3) (78, 3) (79, 3) (80, 3) (81, 0) (82, 0) (83, 0) (84, 0) (85, 0) (86, 0) (87, 0) (88, 0) (89, 0) (90, 0) (91, 0) (92, 0) (93, 0) (94, 0) (95, 0) (96, 0) (97, 0) (98, 0) (99, 0) (100, 0) (101, 0) (102, 0) (103, 0) (104, 0) (105, 0) (106, 0) (107, 0) (108, 0) (109, 0) (110, 0) (111, 0) (112, 0) (113, 0) (114, 0) (115, 0) (116, 0) (117, 0) (118, 0) (119, 0) (120, 0) (121, 0) (122, 0) (123, 0) (124, 0) (125, 0) (126, 0) (127, 0) (128, 0) (129, 0) (130, 0) (131, 0) (132, 0) (133, 0) (134, 0) (135, 0) (136, 0) (137, 0) (138, 0) (139, 0) (140, 0) (141, 0) (142, 0) (143, 0) (144, 0) (145, 0) (146, 0) (147, 0) (148, 0) (149, 0) (150, 0) (151, 0) (152, 0) (153, 0) (154, 0) (155, 0) (156, 0) (157, 0) (158, 0) (159, 0) (160, 0) (161, 0) (162, 0) (163, 0) (164, 0) (165, 0) (166, 0) (167, 0) (168, 0) (169, 0) (170, 0) (171, 0) (172, 0) (173, 0) (174, 0) (175, 0) (176, 0) (177, 0) (178, 0) (179, 0) (180, 0) (181, 0) (182, 0) (183, 0) (184, 0) (185, 0) (186, 0) (187, 0) (188, 0) (189, 0) (190, 0) (191, 0) (192, 0) (193, 0) (194, 0) (195, 0) (196, 0) (197, 0) (198, 0) (199, 0) (200, 0) (201, 0) (202, 0) (203, 0) (204, 0) (205, 0) (206, 0) (207, 0) (208, 0) (209, 0) (210, 0) (211, 0) (212, 0) (213, 0) (214, 0) (215, 0) (216, 0) (217, 0) (218, 0) (219, 0) (220, 0) (221, 0) (222, 0) (223, 0) (224, 0) (225, 0) (226, 0) (227, 0) (228, 0) (229, 0) (230, 0) (231, 0) (232, 0) (233, 0) (234, 0) (235, 0) (236, 0) (237, 0) (238, 0) (239, 0) (240, 0) (241, 4) (242, 0) (243, 0) (244, 0) (245, 0) (246, 0) (247, 0) (248, 0) (249, 0) (250, 0) (251, 0) (252, 0) (253, 0) (254, 0) (255, 0) (256, 0) (257, 0) (258, 0) (259, 0) (260, 0) (261, 0) (262, 0) (263, 0) (264, 0) (265, 0) (266, 0) (267, 0) (268, 0) (269, 0) (270, 0) (271, 0) (272, 0) (273, 0) (274, 0) (275, 0) (276, 0) (277, 0) (278, 0) (279, 0) (280, 0) (281, 0) (282, 0) (283, 0) (284, 0) (285, 0) (286, 0) (287, 0) (288, 0) (289, 0) (290, 0) (291, 0) (292, 0) (293, 0) (294, 0) (295, 0) (296, 0) (297, 0) (298, 0) (299, 0) (300, 0) (301, 0) (302, 0) (303, 0) (304, 0) (305, 0) (306, 0) (307, 0) (308, 0) (309, 0) (310, 0) (311, 0) (312, 0) (313, 0) (314, 0) (315, 0) (316, 0) (317, 0) (318, 1) (319, 1) (320, 1) (321, 1) (322, 1) (323, 1) (324, 1) (325, 1) (326, 1) (327, 1) (328, 1) (329, 1) (330, 1) (331, 1) (332, 0) (333, 0) (334, 0) (335, 0) (336, 0) (337, 0) (338, 0) (339, 0) (340, 0) (341, 0) (342, 0) (343, 0) (344, 0) (345, 0) (346, 0) (347, 0) (348, 0) (349, 0) (350, 0) (351, 0) (352, 0) (353, 0) (354, 0) (355, 0) (356, 0) (357, 0) (358, 0) (359, 0) (360, 0) (361, 0) (362, 0) (363, 0) (364, 0) (365, 0) (366, 0) (367, 0) (368, 0) (369, 0) (370, 0) (371, 0) (372, 0) (373, 0) (374, 0) (375, 0) (376, 0) (377, 0) (378, 0) (379, 0) (380, 0) (381, 0) (382, 0) (383, 0) (384, 0) (385, 0) (386, 0) (387, 0) (388, 0) (389, 0) (390, 0) (391, 0) (392, 0) (393, 0) (394, 0) (395, 0) (396, 0) (397, 0) (398, 0) (399, 0) (400, 0) (401, 0) (402, 0) (403, 0) (404, 0) (405, 0) (406, 0) (407, 0) (408, 0) (409, 0) (410, 0) (411, 0) (412, 0) (413, 0) (414, 0) (415, 0) (416, 0) (417, 0) (418, 0) (419, 0) (420, 0) (421, 0) (422, 0) (423, 0) (424, 0) (425, 0) (426, 0) (427, 0) (428, 0) (429, 0) (430, 0) (431, 0) (432, 0) (433, 0) (434, 0) (435, 0) (436, 0) (437, 0) (438, 0) (439, 0) (440, 0) (441, 0) (442, 0) (443, 0) (444, 0) (445, 0) (446, 0) (447, 0) (448, 0) (449, 0) (450, 0) (451, 0) (452, 0) (453, 0) (454, 0) (455, 0) (456, 0) (457, 0) (458, 0) (459, 0) (460, 0) (461, 0) (462, 0) (463, 0) (464, 0) (465, 0) (466, 0) (467, 0) (468, 0) (469, 0) (470, 0) (471, 0) (472, 0) (473, 0) (474, 0) (475, 0) (476, 0) (477, 0) (478, 0) (479, 0) (480, 0) (481, 0) (482, 0) (483, 0) (484, 0) (485, 0) (486, 0) (487, 0) (488, 0) (489, 0) (490, 0) (491, 0) (492, 0) (493, 0) (494, 0) (495, 0) (496, 0) (497, 0) (498, 0) (499, 0) (500, 0) (501, 0) (502, 0) (503, 0) (504, 0) (505, 0) (506, 0) (507, 0) (508, 0) (509, 0) (510, 0) (511, 0) (512, 0) (513, 0) (514, 0) (515, 0) (516, 0) (517, 0) (518, 0) (519, 0) (520, 0) (521, 0) (522, 0) (523, 0) (524, 0) (525, 0) (526, 0) (527, 0) (528, 0) (529, 0) (530, 0) (531, 0) (532, 0) (533, 0) (534, 0) (535, 0) (536, 0) (537, 0) (538, 0) (539, 0) (540, 0) (541, 0) (542, 0) (543, 0) (544, 0) (545, 0) (546, 0) (547, 0) (548, 0) (549, 0) (550, 0) (551, 0) (552, 0) (553, 0) (554, 0) (555, 0) (556, 0) (557, 0) (558, 0) (559, 0) (560, 0) (561, 0) (562, 0) (563, 0) (564, 0) (565, 0) (566, 0) (567, 0) (568, 0) (569, 0) (570, 0) (571, 0) (572, 0) (573, 0) (574, 0) (575, 0) (576, 0) (577, 0) (578, 0) (579, 0) (580, 0) (581, 0) (582, 0) (583, 0) (584, 0) (585, 0) (586, 0) (587, 0) (588, 0) (589, 0) (590, 0) (591, 0) (592, 0) (593, 0) (594, 0) (595, 0) (596, 0) (597, 0) (598, 0) (599, 0) (600, 4) (601, 0) (602, 0) (603, 0) (604, 0) (605, 0) (606, 0) (607, 0) (608, 0) (609, 0) (610, 0) (611, 0) (612, 0) (613, 0) (614, 0) (615, 0) (616, 0) (617, 0) (618, 0) (619, 0) (620, 0) (621, 0) (622, 0) (623, 0) (624, 0) (625, 0) (626, 0) (627, 0) (628, 0) (629, 0) (630, 0) (631, 0) (632, 0) (633, 0) (634, 0) (635, 0) (636, 0) (637, 0) (638, 0) (639, 0) (640, 0) (641, 0) (642, 0) (643, 0) (644, 0) (645, 0) (646, 0) (647, 0) (648, 0) (649, 0) (650, 0) (651, 0) (652, 0) (653, 0) (654, 0) (655, 0) (656, 0) (657, 0) (658, 0) (659, 0) (660, 0) (661, 0) (662, 0) (663, 0) (664, 0) (665, 0) (666, 0) (667, 0) (668, 0) (669, 0) (670, 0) (671, 0) (672, 0) (673, 0) (674, 0) (675, 0) (676, 0) (677, 0) (678, 0) (679, 0) (680, 0) (681, 0) (682, 0) (683, 0) (684, 0) (685, 0) (686, 0) (687, 0) (688, 0) (689, 0) (690, 0) (691, 0) (692, 0) (693, 0) (694, 0) (695, 0) (696, 0) (697, 0) (698, 0) (699, 0) (700, 0) (701, 0) (702, 0) (703, 0) (704, 0) (705, 0) (706, 0) (707, 0) (708, 0) (709, 0) (710, 0) (711, 0) (712, 0) (713, 0) (714, 0) (715, 0) (716, 0) (717, 0) (718, 0) (719, 0) (720, 0) (721, 0) (722, 0) (723, 0) (724, 0) (725, 0) (726, 0) (727, 0) (728, 0) (729, 0) (730, 0) (731, 0) (732, 0) (733, 0) (734, 0) (735, 0) (736, 0) (737, 0) (738, 0) (739, 0) (740, 0) (741, 0) (742, 0) (743, 0) (744, 0) (745, 0) (746, 0) (747, 0) (748, 0) (749, 0) (750, 0) (751, 0) (752, 0) (753, 0) (754, 0) (755, 0) (756, 0) (757, 0) (758, 0) (759, 0) (760, 0) (761, 0) (762, 0) (763, 0) (764, 0) (765, 0) (766, 0) (767, 0) (768, 0) (769, 0) (770, 0) (771, 0) (772, 0) (773, 0) (774, 0) (775, 0) (776, 0) (777, 0) (778, 0) (779, 0) (780, 0) (781, 0) (782, 0) (783, 0) (784, 0) (785, 0) (786, 0) (787, 0) (788, 0) (789, 0) (790, 0) (791, 0) (792, 0) (793, 0) (794, 0) (795, 0) (796, 0) (797, 0) (798, 0) (799, 0) (800, 0) (801, 0) (802, 0) (803, 0) (804, 0) (805, 0) (806, 0) (807, 0) (808, 0) (809, 0) (810, 0) (811, 0) (812, 0) (813, 0) (814, 0) (815, 0) (816, 0) (817, 0) (818, 0) (819, 0) (820, 0) (821, 0) (822, 0) (823, 0) (824, 0) (825, 0) (826, 0) (827, 2) (828, 2) (829, 2) (830, 2) (831, 2) (832, 2) (833, 2) (834, 2) (835, 2) (836, 2) (837, 2) (838, 2) (839, 2) (840, 2) (841, 2) (842, 2) (843, 2) (844, 2) (845, 2) (846, 2) (847, 2) (848, 2) (849, 2) (850, 2) (851, 2) (852, 2) (853, 2) (854, 2) (855, 2) (856, 2) (857, 2) (858, 2) (859, 2) (860, 2) (861, 2) (862, 2) (863, 2) (864, 2) (865, 2) (866, 2) (867, 2) (868, 2) (869, 2) (870, 2) (871, 2) (872, 2) (873, 2) (874, 2) (875, 2) (876, 2) (877, 2) (878, 2) (879, 2) (880, 2) (881, 2) (882, 2) (883, 2) (884, 2) (885, 2) (886, 2) (887, 2) (888, 2) (889, 2) (890, 2) (891, 2) (892, 2) (893, 2) (894, 2) (895, 2) (896, 2) (897, 2) (898, 2) (899, 2) (900, 2) (901, 2) (902, 2) (903, 2) (904, 2) (905, 2) (906, 2) (907, 2) (908, 2) (909, 2) (910, 2) (911, 2) (912, 2) (913, 2) (914, 2) (915, 2) (916, 2) (917, 2) (918, 2) (919, 2) (920, 2) (921, 2) (922, 2) (923, 0) (924, 0) (925, 0) (926, 0) (927, 0) (928, 0) (929, 0) (930, 0) (931, 0) (932, 0) (933, 0) (934, 0) (935, 0) (936, 0) (937, 0) (938, 0) (939, 0) (940, 0) (941, 0) (942, 0) (943, 0) (944, 0) (945, 0) (946, 0) (947, 0) (948, 0) (949, 0) (950, 0) (951, 0) (952, 0) (953, 0) (954, 0) (955, 0) (956, 0) (957, 0) (958, 0) (959, 0) (960, 0) (961, 0) (962, 0) (963, 0) (964, 0) (965, 0) (966, 0) (967, 0) (968, 0) (969, 0) (970, 0) (971, 0) (972, 0) (973, 0) (974, 0) (975, 0) (976, 0) (977, 0) (978, 0) (979, 0) (980, 0) (981, 0) (982, 0) (983, 0) (984, 0) (985, 0) (986, 0) (987, 0) (988, 0) (989, 0) (990, 0) (991, 0) (992, 0) (993, 0) (994, 0) (995, 0) (996, 0) (997, 0) (998, 0) (999, 0) (1000, 0) (1001, 0) (1002, 0) (1003, 0) (1004, 0) (1005, 0) (1006, 0) (1007, 0) (1008, 0) (1009, 0) (1010, 0) (1011, 0) (1012, 0) (1013, 0) (1014, 0) (1015, 0) (1016, 0) (1017, 0) (1018, 0) (1019, 0) (1020, 0) (1021, 0) (1022, 0) (1023, 0) (1024, 0) (1025, 0) (1026, 0) (1027, 1) (1028, 1) (1029, 1) (1030, 1) (1031, 1) (1032, 1) (1033, 1) (1034, 1) (1035, 1) (1036, 1) (1037, 1) (1038, 1) (1039, 1) (1040, 1) (1041, 1) (1042, 1) (1043, 1) (1044, 1) (1045, 1) (1046, 1) (1047, 1) (1048, 1) (1049, 1) (1050, 1) (1051, 1) (1052, 1) (1053, 1) (1054, 1)};


    \addplot[
    color=rred,
    mark=none,
    line width=0.1mm,
    ]
    coordinates {(0, 0.0) (1, 0.0) (2, 0.0) (3, 0.0) (4, 0.0) (5, 0.0) (6, 3.0) (7, 3.0) (8, 3.0) (9, 3.0) (10, 3.0) (11, 3.0) (12, 3.0) (13, 3.0) (14, 3.0) (15, 3.0) (16, 3.0) (17, 3.0) (18, 3.0) (19, 3.0) (20, 3.0) (21, 3.0) (22, 3.0) (23, 3.0) (24, 3.0) (25, 3.0) (26, 3.0) (27, 3.0) (28, 3.0) (29, 3.0) (30, 3.0) (31, 3.0) (32, 3.0) (33, 3.0) (34, 3.0) (35, 3.0) (36, 3.0) (37, 3.0) (38, 3.0) (39, 3.0) (40, 3.0) (41, 3.0) (42, 3.0) (43, 3.0) (44, 3.0) (45, 3.0) (46, 3.0) (47, 3.0) (48, 3.0) (49, 3.0) (50, 3.0) (51, 3.0) (52, 3.0) (53, 3.0) (54, 3.0) (55, 3.0) (56, 3.0) (57, 3.0) (58, 3.0) (59, 3.0) (60, 0.0) (61, 0.0) (62, 0.0) (63, 0.0) (64, 0.0) (65, 0.0) (66, 3.0) (67, 3.0) (68, 3.0) (69, 3.0) (70, 3.0) (71, 3.0) (72, 0.0) (73, 0.0) (74, 0.0) (75, 0.0) (76, 0.0) (77, 0.0) (78, 0.0) (79, 0.0) (80, 0.0) (81, 0.0) (82, 0.0) (83, 0.0) (84, 0.0) (85, 0.0) (86, 0.0) (87, 0.0) (88, 0.0) (89, 0.0) (90, 0.0) (91, 0.0) (92, 0.0) (93, 0.0) (94, 0.0) (95, 0.0) (96, 0.0) (97, 0.0) (98, 0.0) (99, 0.0) (100, 0.0) (101, 0.0) (102, 0.0) (103, 0.0) (104, 0.0) (105, 0.0) (106, 0.0) (107, 0.0) (108, 0.0) (109, 0.0) (110, 0.0) (111, 0.0) (112, 0.0) (113, 0.0) (114, 0.0) (115, 0.0) (116, 0.0) (117, 0.0) (118, 0.0) (119, 0.0) (120, 0.0) (121, 0.0) (122, 0.0) (123, 0.0) (124, 0.0) (125, 0.0) (126, 0.0) (127, 0.0) (128, 0.0) (129, 0.0) (130, 0.0) (131, 0.0) (132, 0.0) (133, 0.0) (134, 0.0) (135, 0.0) (136, 0.0) (137, 0.0) (138, 0.0) (139, 0.0) (140, 0.0) (141, 0.0) (142, 0.0) (143, 0.0) (144, 0.0) (145, 0.0) (146, 0.0) (147, 0.0) (148, 0.0) (149, 0.0) (150, 0.0) (151, 0.0) (152, 0.0) (153, 0.0) (154, 0.0) (155, 0.0) (156, 0.0) (157, 0.0) (158, 0.0) (159, 0.0) (160, 0.0) (161, 0.0) (162, 0.0) (163, 0.0) (164, 0.0) (165, 0.0) (166, 0.0) (167, 0.0) (168, 0.0) (169, 0.0) (170, 0.0) (171, 0.0) (172, 0.0) (173, 0.0) (174, 0.0) (175, 0.0) (176, 0.0) (177, 0.0) (178, 0.0) (179, 0.0) (180, 0.0) (181, 0.0) (182, 0.0) (183, 0.0) (184, 0.0) (185, 0.0) (186, 0.0) (187, 0.0) (188, 0.0) (189, 0.0) (190, 0.0) (191, 0.0) (192, 0.0) (193, 0.0) (194, 0.0) (195, 0.0) (196, 0.0) (197, 0.0) (198, 0.0) (199, 0.0) (200, 0.0) (201, 0.0) (202, 0.0) (203, 0.0) (204, 0.0) (205, 0.0) (206, 0.0) (207, 0.0) (208, 0.0) (209, 0.0) (210, 0.0) (211, 0.0) (212, 0.0) (213, 0.0) (214, 0.0) (215, 0.0) (216, 0.0) (217, 0.0) (218, 0.0) (219, 0.0) (220, 0.0) (221, 0.0) (222, 0.0) (223, 0.0) (224, 0.0) (225, 0.0) (226, 0.0) (227, 0.0) (228, 0.0) (229, 0.0) (230, 0.0) (231, 0.0) (232, 0.0) (233, 0.0) (234, 0.0) (235, 0.0) (236, 0.0) (237, 0.0) (238, 0.0) (239, 0.0) (240, 0.0) (241, 4.0) (242, 0.0) (243, 0.0) (244, 0.0) (245, 0.0) (246, 0.0) (247, 0.0) (248, 0.0) (249, 0.0) (250, 0.0) (251, 0.0) (252, 0.0) (253, 0.0) (254, 0.0) (255, 0.0) (256, 0.0) (257, 0.0) (258, 0.0) (259, 0.0) (260, 0.0) (261, 0.0) (262, 0.0) (263, 0.0) (264, 0.0) (265, 0.0) (266, 0.0) (267, 0.0) (268, 0.0) (269, 0.0) (270, 0.0) (271, 0.0) (272, 0.0) (273, 0.0) (274, 0.0) (275, 0.0) (276, 0.0) (277, 0.0) (278, 0.0) (279, 0.0) (280, 0.0) (281, 0.0) (282, 0.0) (283, 0.0) (284, 0.0) (285, 0.0) (286, 0.0) (287, 0.0) (288, 0.0) (289, 0.0) (290, 0.0) (291, 0.0) (292, 0.0) (293, 0.0) (294, 0.0) (295, 0.0) (296, 0.0) (297, 0.0) (298, 0.0) (299, 0.0) (300, 0.0) (301, 0.0) (302, 0.0) (303, 0.0) (304, 0.0) (305, 0.0) (306, 0.0) (307, 0.0) (308, 0.0) (309, 0.0) (310, 0.0) (311, 0.0) (312, 0.0) (313, 0.0) (314, 0.0) (315, 0.0) (316, 0.0) (317, 0.0) (318, 0.0) (319, 0.0) (320, 0.0) (321, 1.0) (322, 1.0) (323, 1.0) (324, 1.0) (325, 1.0) (326, 1.0) (327, 1.0) (328, 1.0) (329, 1.0) (330, 1.0) (331, 1.0) (332, 1.0) (333, 1.0) (334, 0.0) (335, 0.0) (336, 0.0) (337, 0.0) (338, 0.0) (339, 0.0) (340, 0.0) (341, 0.0) (342, 0.0) (343, 0.0) (344, 0.0) (345, 0.0) (346, 0.0) (347, 0.0) (348, 0.0) (349, 0.0) (350, 0.0) (351, 0.0) (352, 0.0) (353, 0.0) (354, 0.0) (355, 0.0) (356, 0.0) (357, 0.0) (358, 0.0) (359, 0.0) (360, 0.0) (361, 0.0) (362, 0.0) (363, 0.0) (364, 0.0) (365, 0.0) (366, 0.0) (367, 0.0) (368, 0.0) (369, 0.0) (370, 0.0) (371, 0.0) (372, 0.0) (373, 0.0) (374, 0.0) (375, 0.0) (376, 0.0) (377, 0.0) (378, 0.0) (379, 0.0) (380, 0.0) (381, 0.0) (382, 0.0) (383, 0.0) (384, 0.0) (385, 3.0) (386, 3.0) (387, 3.0) (388, 3.0) (389, 3.0) (390, 3.0) (391, 3.0) (392, 3.0) (393, 3.0) (394, 3.0) (395, 3.0) (396, 3.0) (397, 3.0) (398, 3.0) (399, 3.0) (400, 3.0) (401, 3.0) (402, 3.0) (403, 3.0) (404, 3.0) (405, 0.0) (406, 0.0) (407, 0.0) (408, 0.0) (409, 0.0) (410, 0.0) (411, 0.0) (412, 0.0) (413, 0.0) (414, 0.0) (415, 0.0) (416, 0.0) (417, 0.0) (418, 0.0) (419, 0.0) (420, 0.0) (421, 0.0) (422, 0.0) (423, 0.0) (424, 0.0) (425, 0.0) (426, 0.0) (427, 0.0) (428, 0.0) (429, 0.0) (430, 0.0) (431, 0.0) (432, 0.0) (433, 0.0) (434, 0.0) (435, 0.0) (436, 0.0) (437, 0.0) (438, 0.0) (439, 0.0) (440, 0.0) (441, 0.0) (442, 0.0) (443, 0.0) (444, 0.0) (445, 0.0) (446, 0.0) (447, 0.0) (448, 0.0) (449, 0.0) (450, 0.0) (451, 0.0) (452, 0.0) (453, 0.0) (454, 0.0) (455, 0.0) (456, 0.0) (457, 0.0) (458, 0.0) (459, 0.0) (460, 0.0) (461, 0.0) (462, 0.0) (463, 0.0) (464, 0.0) (465, 0.0) (466, 0.0) (467, 0.0) (468, 0.0) (469, 0.0) (470, 0.0) (471, 0.0) (472, 0.0) (473, 0.0) (474, 0.0) (475, 0.0) (476, 0.0) (477, 0.0) (478, 0.0) (479, 0.0) (480, 0.0) (481, 0.0) (482, 0.0) (483, 0.0) (484, 0.0) (485, 0.0) (486, 0.0) (487, 0.0) (488, 0.0) (489, 0.0) (490, 0.0) (491, 0.0) (492, 0.0) (493, 0.0) (494, 0.0) (495, 0.0) (496, 0.0) (497, 0.0) (498, 0.0) (499, 0.0) (500, 0.0) (501, 0.0) (502, 0.0) (503, 0.0) (504, 0.0) (505, 0.0) (506, 0.0) (507, 0.0) (508, 0.0) (509, 0.0) (510, 0.0) (511, 0.0) (512, 0.0) (513, 0.0) (514, 0.0) (515, 0.0) (516, 0.0) (517, 0.0) (518, 0.0) (519, 0.0) (520, 0.0) (521, 0.0) (522, 0.0) (523, 0.0) (524, 0.0) (525, 0.0) (526, 0.0) (527, 0.0) (528, 0.0) (529, 0.0) (530, 0.0) (531, 0.0) (532, 0.0) (533, 0.0) (534, 0.0) (535, 0.0) (536, 0.0) (537, 0.0) (538, 0.0) (539, 0.0) (540, 0.0) (541, 0.0) (542, 0.0) (543, 0.0) (544, 0.0) (545, 0.0) (546, 0.0) (547, 0.0) (548, 0.0) (549, 0.0) (550, 0.0) (551, 0.0) (552, 0.0) (553, 0.0) (554, 0.0) (555, 0.0) (556, 0.0) (557, 0.0) (558, 0.0) (559, 0.0) (560, 0.0) (561, 0.0) (562, 0.0) (563, 0.0) (564, 0.0) (565, 0.0) (566, 0.0) (567, 0.0) (568, 0.0) (569, 0.0) (570, 0.0) (571, 0.0) (572, 0.0) (573, 0.0) (574, 0.0) (575, 0.0) (576, 0.0) (577, 0.0) (578, 0.0) (579, 0.0) (580, 0.0) (581, 0.0) (582, 0.0) (583, 0.0) (584, 0.0) (585, 0.0) (586, 0.0) (587, 0.0) (588, 0.0) (589, 0.0) (590, 0.0) (591, 0.0) (592, 0.0) (593, 0.0) (594, 0.0) (595, 0.0) (596, 0.0) (597, 0.0) (598, 0.0) (599, 0.0) (600, 4.0) (601, 0.0) (602, 0.0) (603, 0.0) (604, 0.0) (605, 0.0) (606, 0.0) (607, 0.0) (608, 0.0) (609, 0.0) (610, 0.0) (611, 0.0) (612, 0.0) (613, 0.0) (614, 0.0) (615, 0.0) (616, 0.0) (617, 0.0) (618, 0.0) (619, 0.0) (620, 0.0) (621, 0.0) (622, 0.0) (623, 0.0) (624, 0.0) (625, 0.0) (626, 0.0) (627, 0.0) (628, 0.0) (629, 0.0) (630, 0.0) (631, 0.0) (632, 0.0) (633, 0.0) (634, 0.0) (635, 0.0) (636, 0.0) (637, 0.0) (638, 0.0) (639, 0.0) (640, 0.0) (641, 0.0) (642, 0.0) (643, 0.0) (644, 0.0) (645, 0.0) (646, 0.0) (647, 0.0) (648, 0.0) (649, 0.0) (650, 0.0) (651, 0.0) (652, 0.0) (653, 0.0) (654, 0.0) (655, 0.0) (656, 0.0) (657, 0.0) (658, 0.0) (659, 0.0) (660, 0.0) (661, 0.0) (662, 0.0) (663, 0.0) (664, 0.0) (665, 0.0) (666, 0.0) (667, 0.0) (668, 0.0) (669, 0.0) (670, 0.0) (671, 0.0) (672, 0.0) (673, 0.0) (674, 0.0) (675, 0.0) (676, 0.0) (677, 0.0) (678, 0.0) (679, 0.0) (680, 0.0) (681, 0.0) (682, 0.0) (683, 0.0) (684, 3.0) (685, 3.0) (686, 3.0) (687, 3.0) (688, 3.0) (689, 3.0) (690, 0.0) (691, 0.0) (692, 0.0) (693, 0.0) (694, 0.0) (695, 0.0) (696, 0.0) (697, 0.0) (698, 0.0) (699, 0.0) (700, 0.0) (701, 0.0) (702, 0.0) (703, 0.0) (704, 0.0) (705, 0.0) (706, 0.0) (707, 0.0) (708, 0.0) (709, 0.0) (710, 0.0) (711, 0.0) (712, 0.0) (713, 0.0) (714, 0.0) (715, 0.0) (716, 0.0) (717, 0.0) (718, 0.0) (719, 0.0) (720, 0.0) (721, 0.0) (722, 0.0) (723, 0.0) (724, 0.0) (725, 0.0) (726, 0.0) (727, 0.0) (728, 0.0) (729, 0.0) (730, 0.0) (731, 0.0) (732, 0.0) (733, 0.0) (734, 0.0) (735, 0.0) (736, 0.0) (737, 0.0) (738, 0.0) (739, 0.0) (740, 0.0) (741, 0.0) (742, 0.0) (743, 0.0) (744, 0.0) (745, 0.0) (746, 0.0) (747, 0.0) (748, 0.0) (749, 0.0) (750, 0.0) (751, 0.0) (752, 0.0) (753, 0.0) (754, 0.0) (755, 0.0) (756, 0.0) (757, 0.0) (758, 0.0) (759, 0.0) (760, 0.0) (761, 0.0) (762, 0.0) (763, 0.0) (764, 0.0) (765, 0.0) (766, 0.0) (767, 0.0) (768, 0.0) (769, 0.0) (770, 0.0) (771, 0.0) (772, 0.0) (773, 0.0) (774, 0.0) (775, 0.0) (776, 0.0) (777, 0.0) (778, 0.0) (779, 0.0) (780, 0.0) (781, 0.0) (782, 0.0) (783, 0.0) (784, 0.0) (785, 0.0) (786, 0.0) (787, 0.0) (788, 0.0) (789, 0.0) (790, 0.0) (791, 0.0) (792, 0.0) (793, 0.0) (794, 0.0) (795, 0.0) (796, 0.0) (797, 0.0) (798, 0.0) (799, 0.0) (800, 0.0) (801, 0.0) (802, 0.0) (803, 0.0) (804, 0.0) (805, 0.0) (806, 0.0) (807, 0.0) (808, 0.0) (809, 0.0) (810, 0.0) (811, 0.0) (812, 0.0) (813, 0.0) (814, 0.0) (815, 0.0) (816, 0.0) (817, 0.0) (818, 0.0) (819, 0.0) (820, 0.0) (821, 0.0) (822, 0.0) (823, 0.0) (824, 0.0) (825, 0.0) (826, 0.0) (827, 0.0) (828, 0.0) (829, 0.0) (830, 2.0) (831, 2.0) (832, 2.0) (833, 2.0) (834, 2.0) (835, 2.0) (836, 2.0) (837, 2.0) (838, 2.0) (839, 2.0) (840, 2.0) (841, 2.0) (842, 2.0) (843, 2.0) (844, 2.0) (845, 2.0) (846, 2.0) (847, 2.0) (848, 2.0) (849, 2.0) (850, 2.0) (851, 2.0) (852, 2.0) (853, 2.0) (854, 2.0) (855, 2.0) (856, 2.0) (857, 2.0) (858, 2.0) (859, 2.0) (860, 2.0) (861, 2.0) (862, 2.0) (863, 2.0) (864, 2.0) (865, 2.0) (866, 2.0) (867, 2.0) (868, 2.0) (869, 2.0) (870, 2.0) (871, 2.0) (872, 2.0) (873, 2.0) (874, 2.0) (875, 2.0) (876, 2.0) (877, 2.0) (878, 2.0) (879, 2.0) (880, 2.0) (881, 2.0) (882, 2.0) (883, 2.0) (884, 2.0) (885, 2.0) (886, 2.0) (887, 2.0) (888, 2.0) (889, 2.0) (890, 2.0) (891, 2.0) (892, 2.0) (893, 2.0) (894, 2.0) (895, 2.0) (896, 2.0) (897, 2.0) (898, 2.0) (899, 2.0) (900, 2.0) (901, 2.0) (902, 2.0) (903, 2.0) (904, 2.0) (905, 2.0) (906, 2.0) (907, 2.0) (908, 2.0) (909, 2.0) (910, 2.0) (911, 2.0) (912, 2.0) (913, 2.0) (914, 2.0) (915, 2.0) (916, 2.0) (917, 2.0) (918, 2.0) (919, 2.0) (920, 2.0) (921, 0.0) (922, 2.0) (923, 0.0) (924, 0.0) (925, 0.0) (926, 0.0) (927, 0.0) (928, 0.0) (929, 0.0) (930, 0.0) (931, 0.0) (932, 0.0) (933, 0.0) (934, 0.0) (935, 0.0) (936, 0.0) (937, 0.0) (938, 0.0) (939, 0.0) (940, 0.0) (941, 0.0) (942, 0.0) (943, 0.0) (944, 0.0) (945, 0.0) (946, 0.0) (947, 0.0) (948, 0.0) (949, 0.0) (950, 0.0) (951, 0.0) (952, 0.0) (953, 0.0) (954, 0.0) (955, 0.0) (956, 0.0) (957, 0.0) (958, 0.0) (959, 0.0) (960, 0.0) (961, 0.0) (962, 0.0) (963, 0.0) (964, 0.0) (965, 0.0) (966, 0.0) (967, 0.0) (968, 0.0) (969, 0.0) (970, 0.0) (971, 0.0) (972, 0.0) (973, 0.0) (974, 0.0) (975, 0.0) (976, 0.0) (977, 0.0) (978, 0.0) (979, 0.0) (980, 0.0) (981, 0.0) (982, 0.0) (983, 0.0) (984, 0.0) (985, 0.0) (986, 0.0) (987, 0.0) (988, 0.0) (989, 0.0) (990, 0.0) (991, 0.0) (992, 0.0) (993, 0.0) (994, 0.0) (995, 0.0) (996, 0.0) (997, 0.0) (998, 0.0) (999, 0.0) (1000, 0.0) (1001, 0.0) (1002, 0.0) (1003, 0.0) (1004, 0.0) (1005, 0.0) (1006, 0.0) (1007, 0.0) (1008, 0.0) (1009, 0.0) (1010, 0.0) (1011, 0.0) (1012, 0.0) (1013, 0.0) (1014, 0.0) (1015, 0.0) (1016, 0.0) (1017, 0.0) (1018, 0.0) (1019, 0.0) (1020, 0.0) (1021, 0.0) (1022, 0.0) (1023, 0.0) (1024, 0.0) (1025, 0.0) (1026, 0.0) (1027, 0.0) (1028, 0.0) (1029, 0.0) (1030, 1.0) (1031, 1.0) (1032, 1.0) (1033, 1.0) (1034, 1.0) (1035, 1.0) (1036, 1.0) (1037, 1.0) (1038, 1.0) (1039, 1.0) (1040, 1.0) (1041, 1.0) (1042, 1.0) (1043, 1.0) (1044, 1.0) (1045, 1.0) (1046, 1.0) (1047, 1.0) (1048, 1.0) (1049, 1.0) (1050, 1.0) (1051, 1.0) (1052, 1.0) (1053, 1.0) (1054, 1.0)};
    
        

\end{axis}
\end{tikzpicture}
    \caption{Unfiltered (a) and filtered (b) predicted and true test dataset labels.} \label{fig:time_series_plot}

\end{figure}
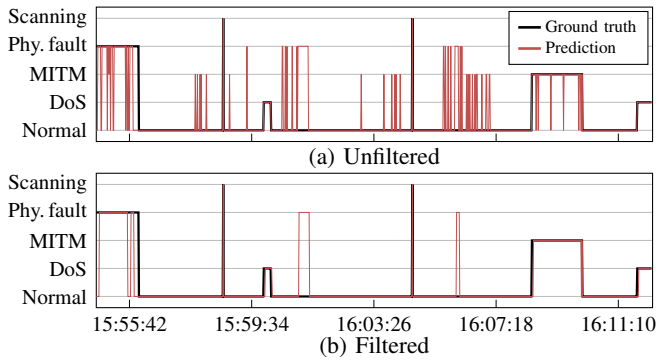

%% file: 05_Conclusion.tex
\section{Conclusion and future work} \label{sec:conclusion}
This work assesses \gls{ML}-based cyber-physical intrusion detection and multi-class classification for \glspl{ICS}. 
For that purpose, a systematic comparison to a purely network data-based approach is conducted, followed by an evaluation of misclassifications and detection delay. 
An average $F^{m}_{1}$ improvement of $15$ percentage points across several supervised classification pipelines demonstrates the benefit of incorporating physical process data into intrusion detection and classification.
Moreover, simultaneous processing of cyber attacks and physical faults is demonstrated, which paves the way to holistic cross-domain root cause analysis.
Based on the evaluation of misclassifications, filtering of the classifier output (prediction filtering) is proposed to reduce false positives, which, however, comes at the cost of higher detection delays.

A remaining problem of cyber-physical intrusion detection and classification is the dependency on usually scarce attack samples. 
A potential solution is seen in applying attack sample-independent unsupervised methods.
The often weaker performance of such methods in distinguishing attack types may be counteracted by considering cyber-physical input data.